\documentclass[12pt]{article}
\usepackage{graphicx} 
\usepackage[margin=1in]{geometry}
\usepackage{amsmath, amssymb}
\usepackage[shortlabels]{enumitem}
\usepackage{amsfonts}
\usepackage{xcolor}
\usepackage{natbib}
\usepackage{dirtytalk}
\usepackage{setspace}
\usepackage{booktabs}
\usepackage{threeparttable}
\newtheorem{theorem}{Theorem}

\newtheorem{algorithm}{Algorithm}

\newtheorem{corollary}{Corollary}

\newtheorem{definition}{Definition}
\newtheorem{example}{Example}

\newtheorem{lemma}{Lemma}

\newtheorem{proposition}{Proposition}

\newtheorem{assumption}{Assumption}

\newcommand{\figurenote}[1]{%
  \par\smallskip
  \begin{minipage}{0.94\linewidth}
    \footnotesize\raggedright
    \textit{Notes:} #1
  \end{minipage}%
}

\title{\vspace{-2cm} Endogenous Identity in a Social Network}
\author{\large Christian Ghiglino and Nicole Tabasso$^{\text{*}}$ }

\date{October 2026}

\begin{document}

\maketitle
\begin{center}

Abstract

\end{center}

We study the interaction between endogenous identity choice and peer influence in social networks. Individuals choose identities that determine behavioral norms and reference groups within a fixed network. Equilibrium peer behavior, in turn, shapes identity choice. Interventions affect behavior both within identities through conformism pressure and through identity choices. Changes that raise actions holding identities fixed can induce identity changes that lower them. Our framework explains why peer effects estimated under one composition may fail to predict the effects of reassignment. It also shows how network structure can sustain heterogeneous behavior among individuals with identical abilities.
\noindent

\vspace*{1.0cm}
\vspace*{0.5cm}

\noindent *{\footnotesize \textit{Ghiglino}: Department of Economics, University of Essex; cghig@essex.ac.uk \\ \textit{Tabasso}: Department of Economics,  Ca’ Foscari University of Venice; nicole.tabasso@unive.it. Tabasso gratefully acknowledges funding from the European Union Next-GenerationEU - National Recovery and Resilience Plan (NRRP) – Mission 4 Component 2, Investment N.1.2 – CUP N.H73C22001340001. The work reflects only the authors’ views and opinions, neither the European Union nor the European Commission can be  considered responsible for them. We are very grateful for helpful comments and discussions from Matt Jackson, Alastair Langtry, Ankur Mani, and seminar participants at the Essex SENA workshop 2024, the $9^{th}$ Annual Conference on Networks Science and Economics, BSE 2024.}\\

\setstretch{1.25}

\section{Introduction}
This paper studies the interaction between endogenous identity choice and peer influence. Both are important determinants of economic behavior. Identities shape the norms individuals seek to follow, while social connections influence behavior through conformity, social comparison, and spillovers. These forces can help explain why individuals facing similar economic opportunities make different choices. They are also interdependent: identity shapes whose behavior individuals regard as a reference point, while their social connections influence which groups they identify with.

Understanding this interdependence matters for policies that seek to improve outcomes by changing social exposure. Economic mobility provides one motivation. \citet{chetty_nature1} show that economic connectedness—the prevalence of friendships between individuals of low and high socioeconomic status—is among the strongest predictors of upward economic mobility. Their work does not identify a unique mechanism through which economic connectedness might affect economic mobility. We formalize one possible channel, which is that these connections change the groups individuals identify with and, consequently, the norms and peer behavior that guide their choices.

The difficulty of designing beneficial social exposure is illustrated by \citet{carrell2013}. They used peer effects estimated from randomly assigned peer groups to construct new groups intended to improve the performance of low-achieving students. The intervention instead reduced their performance. The authors provide evidence that students formed more homogeneous subgroups within the assigned groups, limiting the interactions the intervention sought to encourage. Their findings highlight the importance of distinguishing assigned peer groups from the groups through which peer influence actually operates.

We model the interdependence of identity and social environment. Individuals are embedded in a fixed social network, but choose an identity that determines both the behavioral norm they face and which neighbors constitute their reference group. The attractiveness of an identity depends not only on its status and local social support, but also on the costs of conforming to its members’ equilibrium behavior. An identity associated with higher actions of its members may therefore be less attractive to individuals whose own characteristics make those actions costly to match. Identity choice thus shapes the effective network of peer influence, while anticipated peer behavior feeds back into identity choice.

Our general analysis characterizes this interaction in a heterogeneous population. Every identity profile induces a unique equilibrium in actions, allowing identity choice to be represented by a threshold condition that trades off differences in local social support against the remaining equilibrium payoff differences between identities. Unlike in standard threshold models, these thresholds are endogenous: they depend on the equilibrium behavior generated within the relevant identity groups. Numerical examples show that this feedback can overturn familiar comparative statics: an additional adopter can induce others to leave the identity; a link that raises actions conditional on identities can lower an individual's equilibrium action; and greater homophily along identity and ability lines can induce identity change.  

A homogeneous-ability benchmark then isolates the role of social attachment. Even when individuals are otherwise identical and evaluate each identity identically absent local social support, network structure can sustain different identities and, through their different prescriptions, persistent differences in behavior. In particular, an identity that is otherwise less attractive can survive when its members are sufficiently embedded among one another. Following an increase in the attractiveness of the alternative, lower-degree individuals are the first to become willing to change their identities, while the relative embeddedness of those remaining determines how far identity change spreads. Such diffusion need not improve welfare: gains from adopting the more attractive identity can be outweighed by the loss of same-identity benefits on links crossing the resulting identity boundary.

The model consequently distinguishes two margins through which social exposure affects behavior: peer composition can change behavior within an existing identity, or it can change identity itself and thereby change both the norm and the relevant peer group. These margins provide a common mechanism for several empirical patterns that are usually studied separately. \citet{chetty_nature1} document the importance of cross-class friendships for economic mobility; in our framework, such connections can change the social support for alternative educational identities. \citet{CornoLaFerraraBurns2022} find effects of interracial roommate assignment even after controlling for prior achievement; our model allows equally able peers to exert different influence, depending on which identity and norms they associate themselves with. And the designed peer groups of \citet{carrell2013} illustrate why peer effects estimated under one composition need not predict the consequences of another when changes in composition alter effective reference groups and the incentives to adopt different identities. These studies do not identify the exact mechanism(s) underlying their results; our model provides a single channel through which their different patterns can arise.

This perspective has implications for interventions that seek to exploit peer effects or economic connectedness. Network structure and observable peer characteristics may be insufficient for predicting the effects of reassignment when the choice of identities and reference groups respond endogenously. Information about individuals' identities, perceived behavioral norms, and reference groups may therefore help inform the design and evaluation of such interventions. Moreover, greater exposure to peers with characteristics associated with desirable outcomes need not generate stronger spillovers if the resulting interaction environment makes the associated identity costly to adopt.

Our paper builds most directly on the economic literature on social identity. \citet{akerlof_kranton} formalize the idea that social categories carry prescriptions for behavior, and \citet{akerlofkranton2002} study how students' choices of identity and educational effort depend on their school environment. \citet{shayo2009} develops a framework in which identification and behavior are jointly determined, with identification shaped by group status and perceived distance from the group. Subsequent work has applied and extended endogenous social identification in a range of settings \citep{lindqvist2013identity,grossman2021identity,gennaioli2019identity}, while \citet{klor2010social} provides experimental evidence on the behavioral importance of social identity. Most directly for our mechanism, \citet{atkin_etal_food} show that identification responds to the economic costs of following identity-specific prescriptions. Our model examines how these
behavioral costs are generated by social interactions. Individuals respond both to an identity's prescribed action and to the equilibrium behavior of same-identity neighbors; because identity determines which neighbors constitute the relevant reference group, peer behavior in turn affects identity choice.

Conditional on identities, our action game belongs to the literature on local social comparisons and peer effects in networks
\citep[see, e.g.,][]{ghiglino2010keeping,bramoulle2020survey, Zenou2026PeerNetwork}.
Individuals respond to the average rather than the sum of their same-identity neighbors' actions, as in local-average models
\citep{LiuPatacchiniZenou2014,ushchev_zenou}. Also related,
\citet{ghiglino_langtry} combine social identity with behavioral comparisons among same-identity neighbors, taking identities as given while allowing behavioral benchmarks to be endogenous. We allow individuals to change identity in response to the equilibrium behavior associated with each available identity. This introduces a feedback from peer behavior to the identity choices that determine who influences whom. Consequently, an intervention that raises actions within an identity can induce some members to leave it and reduce their actions. More generally, it implies that changes in the social environment can induce discrete changes in behavior. This offers a source of nonlinear responses to peer behavior complementary to the one emphasized in \citet{BoucherRendallUshchevZenou2024}.  

The endogenous determination of socially relevant peers connects our paper to work on peer-group and network formation, as well as to the literature on cultural assimilation. \citet{BaccaraYariv2013} study endogenous peer-group formation when group composition affects subsequent contributions, while \citet{Hiller2017,Badev2021,SadlerGolub2026} analyze settings in which behavior and social relationships are jointly determined. Relatedly, cultural assimilation and identity respond to the surrounding social environment \citep{BisinPatacchiniVerdierZenou2016,sato2020,Itoh2024}. Most closely, \citet{verdier} study a two-stage setting in which minority individuals choose whether to assimilate to a majority before individuals choose effort in a social network, while \citet{Olcina2024} study integration dynamics when individuals conform to the integration behavior of their network neighbors. In our model, identity choice instead determines which existing social relationships become behaviorally relevant: only same-identity neighbors enter an individual's action reference group. Identity therefore jointly determines the behavioral norm and the set of peers whose equilibrium behavior matters, while that behavior in turn affects the attractiveness of the identity. This mechanism can make changes in peer composition or network structure that increase direct social support for an identity nevertheless discourage its adoption. The fact that these choices operate on an exogenously given network allows us to study how its characteristics affect equilibrium behavior.

The homogeneous-ability benchmark connects to threshold models of collective behavior and network coordination
\citep{granovetter_thresholds,morris_contagion,gagnon-goyal,langtry2024thresholds}. With two identities, identity choice reduces to a threshold rule in relative local social support. This recovers the familiar role of cohesive groups in sustaining different choices and isolates how social attachment can sustain behavioral differences among otherwise identical individuals. With heterogeneous abilities, however, changes in group composition also change the costs of conforming to peers. This distinction generates non-monotone adoption under heterogeneous abilities, while the homogeneous benchmark isolates how network structure alone can sustain different identities and actions among otherwise identical individuals.

The remainder of the paper is organized as follows. Section~\ref{sec:model} introduces the model and Section~\ref{sec:equilibrium} characterizes equilibrium behavior and identity choice. Section~\ref{sec:identity} studies how identity choices depend on individual characteristics and network structure, while Section~\ref{sec:homogeneity} examines identity persistence, diffusion, and welfare in the homogeneous benchmark. Section~\ref{sec:applications} relates our framework to empirical findings on peer effects and economic mobility, and Section~\ref{sec:conclusions} concludes. All proofs and verification of examples are in the Appendix.

\section{The Model} \label{sec:model}
\subsection{Network and Identities}

\textbf{Exogenous Network.}
We model the society as a population \(N = \{1, \dots, n\}\), with $n\geq 2$, of individuals connected through an undirected network $\mathbf{g}$ with adjacency matrix
\(\mathbf{G} = [g_{ij}]\), where \(g_{ij} \in \{0,1\}\), \(g_{ij} = g_{ji}\), and \(g_{ii}=0\). 
Each individual \(i \in N\) has degree \(d_i = \sum_j g_{ij}\) and the network is connected. The network is fixed and common knowledge. To simplify notation, we suppress the dependence of equilibrium objects on $\mathbf{g}$ unless comparative statics with respect to the network are considered.
\smallskip

\noindent \textbf{Identities and Group Structure.} 
Each individual chooses one of \(M \ge 2\) possible \emph{identities}, indexed by the set \(\mathcal{I} = \{1, \dots, M\}\). 
Let $s_i \in \mathcal{I}$ be the identity chosen by individual $i$ and let $s=(s_1,...,s_n)$ denote the identity profile in society. With some abuse of notation, we will write $s=(s_i,s_{-i})$ for the identity profile when individual $i$ chooses identity $s_i$. For each identity $I\in\mathcal{I}$, define $N_I = \{\, i \in N : s_i = I \,\}$ as the subset of individuals choosing identity \(I\). The sets $(N_I)_{I\in\mathcal{I}}$ are pairwise disjoint and their union is $N$; the nonempty sets among them form the identity partition induced by $s$. The size of each set is \(n_I = |N_I|\).

For each individual $i \in N$ and $I \in \mathcal{I}$ let $d_{i,I}$ denote the number of $i$'s neighbors who choose identity $I$. Hence, $i$'s total degree satisfies $d_i = \sum_{I \in \mathcal{I}} d_{i,I}$. Only the degree corresponding to $i$'s own identity, $d_{i,s_i}$, affects their utility at a given profile of identities $s$. The remaining $d_{i,J}$ for $J\neq s_i$ are defined for notational convenience when considering deviations.

For each identity \(I \in \mathcal{I}\), define the \emph{within-identity adjacency matrix} 
\[
\mathbf{G_I}=[g_{ij}]_{ij\in N_I}
\]
as the $n_I \times n_I$ adjacency matrix induced by $N_I$,
that is, the restriction of the full network to links between individuals who share identity \(I\). Similarly to degree, only these within-identity connections matter for social influence. It is analytically convenient to define also the row-normalised $(n_I\times n_I)$-matrix $\mathbf{\hat{G}_I}=[\hat{g}_{ij,I}]$ with entries $\hat{g}_{ij,I}=g_{ij,I}/d_{i,I}$ whenever $d_{i,I}>0$ and $0$ otherwise.

\subsection{Individuals and Preferences}
\textbf{Individuals.}
Each individual $i$ is described by their social identity $s_i\in \mathcal{I}$, their ability $w_i \in \mathbb{R}_{++}$, their position in the network $\mathbf{g}$, and their action level $x_{i} \in \mathbb{R}_{+}$.  Out of these variables, their position in the network $\mathbf{g}$ and their ability are exogenously given, while they can choose their identity and action level. The network structure, abilities, and the functional form of preferences are common knowledge. 
Each identity $I$ confers upon its members a status $\mu_I\geq 0$ and is associated with a \textit{prescribed} action level $v_I\geq 0$. 
\smallskip


\noindent \textbf{Preferences.}
Formally, the utility of individual $i$, choosing identity $s_i$ and action $x_{i}$, when faced with network $\mathbf{g}$ and identity profile $s_{-i}$ 
is given by 
\begin{small}
\begin{equation}
U_i(x_{i}, x_{-i},s) = \mu_{s_i} + \beta d_{i,s_i} + x_{i} - \frac{1}{2w_i}x_{i}^{2} - \frac{\alpha}{2} 1\{d_{i,s_i}>0\}(x_{i} - \Bar{x}_{i})^2 - \frac{\gamma}{2}(x_{i}-v_{s_i})^2.
\label{eq:utility}
\end{equation}
\end{small}

where $\beta,\alpha,\gamma> 0$ are parameters that measure the importance of having neighbors of the same identity ($\beta$), of conforming to the average action these neighbors take ($\alpha$)\footnote{We employ the indicator function $1\{\cdot\}$ as peer pressure is only relevant for individuals that have same-identity connections.} and of conforming to the prescribed action of their chosen identity ($\gamma$), relative to the utility they would receive from taking action $x_{i}$ if all conformist and neighborhood considerations were absent. The average action of $i$'s neighbors $j$ who share $s_j=s_i$, $\Bar{x}_{i}$, is given by
\begin{equation}
    \Bar{x}_{i} = \sum_{j\in N_{s_i}}\hat{g}_{ij,s_i}x_{j}.
\end{equation}

Thus, individuals face pressure to conform both to the average neighbor action and the prescribed action of their chosen identity.\footnote{This idea has been introduced in sociology by \cite{cialdini1991norms}, who call the first \say{descriptive} and the second \say{injunctive} norms.} Note that individuals who have no neighbors of the same identity ($d_{i,s_i}=0$) do not face any pressure to conform to their neighbors and their optimal peer-free action level is
\begin{equation} \label{eq:autarky_x*}
   x_{i}^*(s_i)= [1+\gamma v_{s_i}]\frac{w_i}{1+\gamma w_i},
\end{equation}
which is independent of the identity profile of others, $s_{-i}$. Similarly, it is straightforward to show that if conformism pressure was entirely absent ($\alpha=\gamma=0$), optimality would imply that actions are independent of both $s_i$ and $s_{-i}$, i.e., $x_{i}(s)^*=x_{i}^*=w_i$. 
\smallskip

\noindent \textbf{Timing and Information.}
The network $\mathbf{g}$, the set of identities $\mathcal{I}$, individual abilities $(w_i)_{i\in N}$, the preference parameters ($\alpha, \beta, \gamma$), and each identity's prescribed action $v_I$ and status $\mu_I$ are common knowledge. The game unfolds in two stages, non-cooperatively. In stage $1$, individuals simultaneously choose an identity, resulting in identity profile $s$, which is then publicly observed. In stage $2$, individuals simultaneously choose actions given $s$.\footnote{One could alternatively model identity and actions as chosen simultaneously in a one-shot game. We adopt the two-stage formulation as in many environments identities are more persistent and adjust less frequently than individual actions, so actions are typically chosen conditional on a prevailing identity. In addition, the two-stage formulation imposes sequential rationality, requiring that after any realized identity profile actions constitute an equilibrium of the induced action subgame.} Individuals anticipate the equilibrium outcome of the action subgame induced by any identity profile when choosing identities. 

We solve the two-stage game by backward induction, first characterizing equilibrium action choices for a given identity profile $s$, and then analyzing identity choice in the resulting reduced-form game.

\section{Equilibrium} \label{sec:equilibrium}
\subsection{Stage $2$ equilibrium: action choices}
Fix an identity profile $s$ with $s_i=I$. Given identities, individuals simultaneously choose actions to maximize utility, taking the actions of all others as given. Optimality in actions requires that

\begin{equation}
\label{eq:FOC}
x_{i} = \psi_i [1+\gamma v_{I} + \alpha 1\{d_{i,I}>0\} \bar x_{i}],
\end{equation}
where 
\[
\psi_i = \frac{w_i}{1+w_i(\gamma + 1\{d_{i,I}>0\}\alpha)} \in (0, w_i]
\]
captures the sensitivity of $i$’s optimal action to the prescribed and average neighbor actions of their chosen identity $I$. As utility is strictly concave in $x_i$, this first-order condition is necessary and sufficient. Define the weighted adjacency matrix $\tilde{\mathbf G}_I = [\tilde g_{ij,I}]$, where $\tilde g_{ij,I} = \psi_i \hat g_{ij,I}$ measures the effective influence of neighbor $j$ on $i$ within identity $I$. Then, letting $\mathbf{x}_I := (x_{i})_{i\in N_I}^{\!\top}$ and $\boldsymbol{\psi}_I := (\psi_i)_{i\in N_I}^{\!\top}$ denote the vectors of actions and sensitivities of individuals \textit{holding identity~$I$}, the Nash equilibrium of this action subgame can be written as

\begin{equation}
\label{eq:equilibrium_x}
\mathbf{x}_I^* = (1+\gamma v_I)\mathbf{b}_I,
\end{equation}
where 

\begin{equation} \label{eq:centrality_vector}
    \mathbf{b}_I = (\mathbf{I}_{n_I} - \alpha \tilde{\mathbf G}_I)^{-1}\boldsymbol{\psi}_I
\end{equation}
is a Katz--Bonacich centrality vector associated with identity~$I$, with $b_{i,I}$ denoting the centrality of individual $i$ within identity $I$.\footnote{Since the identity profile $s$ is determined endogenously in the first stage of the game, the influence matrix $\tilde{\mathbf G}_I$ and hence the centrality vector $\mathbf b_I$ are themselves endogenous equilibrium objects.} This yields our first result.

\begin{lemma}[Nash equilibrium in actions]
\label{lem:action_equ}
For any identity profile $s$ and any identity $I\in\mathcal I$ such that $N_I\neq\varnothing$, the action subgame induced on the corresponding identity group $N_I$ admits a unique Nash equilibrium, given by
\[
\mathbf x_I^*=(1+\gamma v_I)\mathbf b_I.
\]
\end{lemma}

Decomposability across identities implies that stacking these identity-specific equilibrium vectors across the nonempty identity groups \(\{N_I\}_{I\in\mathcal I}\) yields the unique Nash equilibrium action profile of the full population, denoted \(x^*(s)\).

\subsection{Stage $1$ equilibrium: identity choice}
Lemma~\ref{lem:action_equ} establishes that every identity profile $s=(s_1,...,s_n)$ induces a unique equilibrium of the stage-$2$ action subgame. We therefore solve the game by backward induction.

Let $s=(I,s_{-i})$ be the profile when individual $i$ adopts identity $I$, and let $x=(x_i,x_{-i})$. Then

    \[
        V_{i}(s) \equiv
    U_i\bigl(x_i^*(s),x_{-i}^*(s),s\bigr)
    \]
denotes individual $i$'s equilibrium payoff from choosing identity $I$. Substituting equilibrium actions into utility therefore yields a well-defined reduced-form game of identity choice in which individuals simultaneously choose identities to maximize $V_{i}(s)$.

Any Nash equilibrium of this reduced-form identity game, together with the continuation equilibria characterized in Lemma~\ref{lem:action_equ}, constitutes a subgame-perfect equilibrium of the original two-stage game. Because the reduced-form identity game is finite, a possibly mixed-strategy Nash equilibrium always exists.

In the remainder of the paper, we focus on pure-strategy equilibria in identities and characterize the conditions under which they exist, as well as which identities are chosen in equilibrium.

\section{Identity choice and network structure} \label{sec:identity}
\subsection{Identity choice as a threshold game} \label{sec:identity_choice}
For an individual $i$, choosing identity $I$ rather than $J$ involves a trade-off between two forces. One is the direct benefit of having neighbors who share one's identity, which is increasing in their numbers. The second is the payoff associated with the identity itself, including its status, prescribed norm, and the equilibrium behavior of the corresponding reference group. We denote this an identity's \emph{core value}. To characterize equilibrium identity choices, it is useful to separate these two components. We therefore define the following concepts. 

\begin{definition} \label{def:intrinsic_value}
    Let
    \begin{equation} \label{eq:V-tilde}
    \widetilde{V}_{i}(I,s_{-i})=V_{i}(I,s_{-i})-\beta d_{i,I}
    \end{equation}
    denote the core value of identity $I$ to individual $i$.
\end{definition}

The core value of an identity captures  all determinants of its attractiveness other than the social support for it, i.e., the direct benefit from having same-identity neighbors. It therefore remains endogenous to the identity profile: changing who holds an identity can change its core value by changing the equilibrium behavior associated with choosing it. Each comparison evaluates the action equilibrium under the identity choice being considered, holding other individuals' identities fixed.

\begin{definition} \label{def:C_{i,I}}
    Let $C_i(I,J,s_{-i})\in \mathbb{R}$ be defined as the \emph{pairwise cost} to individual \( i \) of choosing identity \( I \) rather than \( J \):
   
    \begin{equation} \label{eq:C_{i,I}}
        C_i(I,J,s_{-i}) = \frac{1}{\beta} \left[ \widetilde{V}_{i}(J,s_{-i}) - \widetilde{V}_{i}(I,s_{-i}) \right]
    \end{equation}
\end{definition}

The pairwise cost  $C_i(I,J,s_{-i})$ measures the normalized difference in core values between two identities. Whenever  $ C_i(I,J,s_{-i}) \ge 0$, individual \(i\) finds identity \(J\)'s core value (weakly) more attractive, so choosing \(I\) is costly; when $C_i(I,J,s_{-i})<0$, the reverse holds.

Definitions \ref{def:intrinsic_value} and \ref{def:C_{i,I}} allow us to succinctly write the individual best-response conditions underlying equilibrium.

\begin{lemma}[Threshold representation of identity choice] \label{lem:threshold}
Fix an individual $i$ and an identity profile $s_{-i}$. Identity $I\in\mathcal{I}$ is a best response for $i$ if and only if 
\begin{equation}
\label{eq:ind_threshold}
    d_{i,I}-d_{i,J} \ge C_i(I,J,s_{-i})
    \qquad
    \text{for every }J\in\mathcal I,\ J\neq I.
\end{equation}
\end{lemma}

Lemma~\ref{lem:threshold} shows that an individual's identity choice admits a pairwise threshold representation. The left-hand side measures the difference in direct social support for identities \(I\) and \(J\), while \(C_i(I,J,s_{-i})\) captures their difference in core values. Colloquially, individuals may accept to associate themselves with an identity that offers them lower status, or exposes them to higher conformism pressure, for the simple reason of gaining direct social benefits from sharing an identity with more of their neighbors.

The lemma characterizes an individual's optimal identity conditional on the choices of all other individuals. It allows us to formally characterize the conditions under which pure-strategy equilibria in identity choice exist, and how they look like. To do so, we denote by $d_i(S)\equiv \sum_{j\in S}g_{ij}, \ S\subseteq N$ the number of $i$'s neighbors contained in set $S$. In particular, $d_i(N_I)=d_{i,I}$.

\begin{theorem}[{\small Characterization of pure-strategy identity equilibria}]
\label{prop:equilibria_pure_strategy}
\mbox{}\\
A pure-strategy Nash equilibrium in identity choices exists if and only if there exists an identity profile $s$ such that, for every nonempty identity group $N_I$ in the identity partition induced by $s$, every individual $i\in N_I$, and every \(J\in\mathcal{I}, \ J\neq I\),
\begin{equation} \label{eq:pure-strategy-condition}
    d_i(N_I) - d_i(N_J)\geq C_i(I,J,s_{-i}).
\end{equation}
In particular:
\begin{enumerate}[(i)]
    \item \textbf{Monomorphic equilibrium.} A monomorphic equilibrium in which all individuals choose identity \(I\) exists if and only if \(N_I=N\) and 
    \begin{equation} 
        d_i(N)=d_i\geq C_i(I,J,s_{-i}) \qquad \forall\, i\in N,\;\forall\, J\neq I.
    \end{equation}
    \item \textbf{Coexistence equilibrium.} A coexistence equilibrium exists if and only if there exists a partition \(\{N_I\}_{I\in\mathcal{I}}\) satisfying equation \eqref{eq:pure-strategy-condition} that includes at least two nonempty identity groups. 
\end{enumerate}
\end{theorem}

Theorem~\ref{prop:equilibria_pure_strategy} formalizes that multiple identities can coexist when the network is sufficiently cohesive: individuals remain attached to their identity, provided that their within-group connections exceed the support from each alternative identity group by the relevant switching costs. Importantly, these costs are endogenous and depend on the identity profile $s$. 

\begin{figure} [ht]
    \centering
    \includegraphics[width=0.9\linewidth]{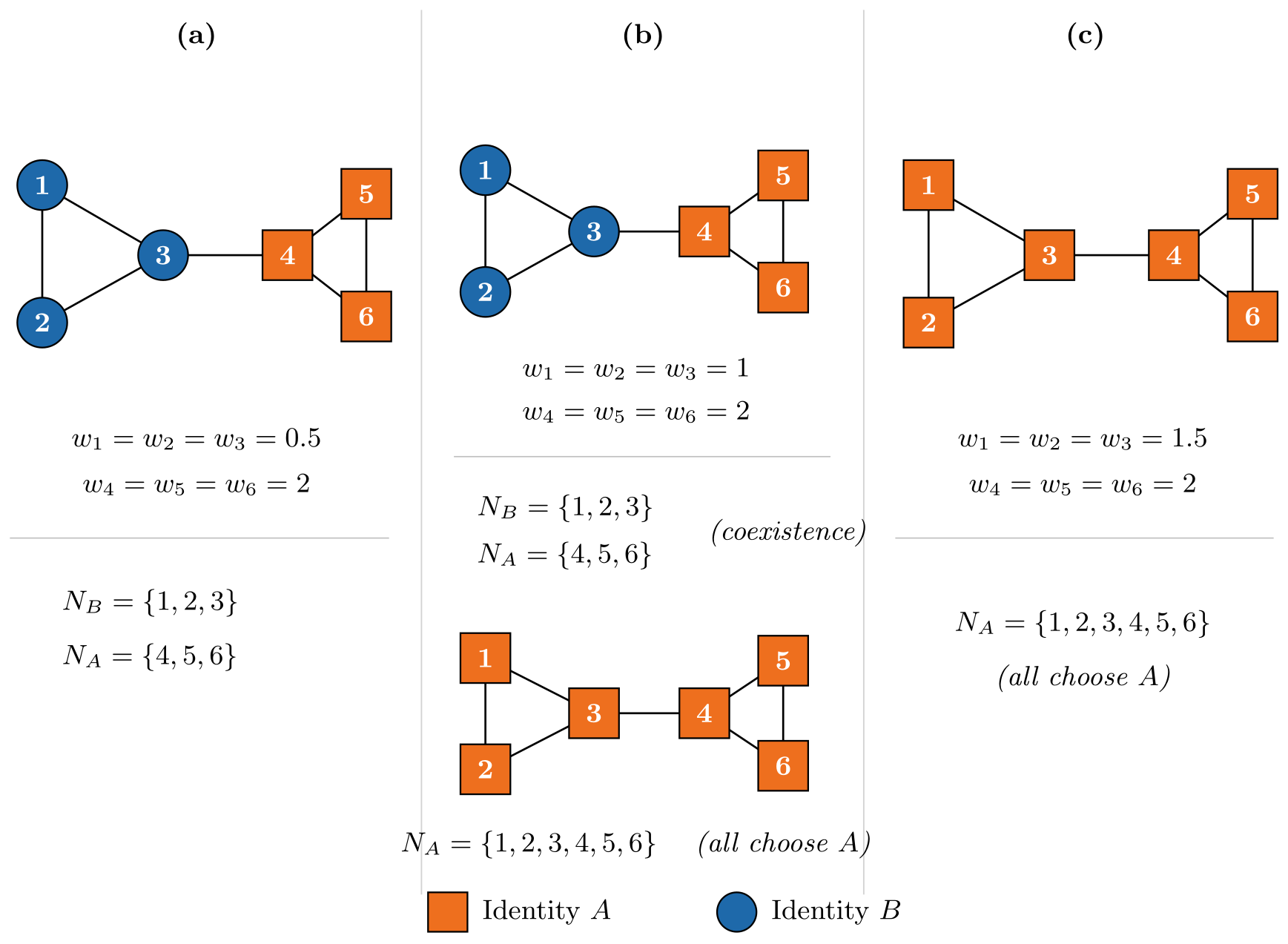}
    \caption{\textbf{Examples of pure-strategy equilibrium structures.}}
    \label{fig:pure-equilibria}
    \figurenote{Orange squares denote \(A\) and blue circles denote \(B\). The figure shows all pure-strategy identity equilibria for the displayed network under the three ability profiles. Parameters are \((\alpha,\beta,\gamma)=(1/4,1/4,3/4)\), \((\mu_A,\mu_B)=(1,0)\), and \((v_A,v_B)=(3,1/2)\).}
\end{figure}

As a consequence, the same exogenous network need not support the same identity structure for populations with different characteristics. Figure \ref{fig:pure-equilibria} illustrates this point for the case of two identities, $\mathcal{I}=\{A,B\}$. Holding the network and all parameters other than abilities fixed, increasing the abilities of individuals \(1,2,3\) changes the set of pure-strategy identity equilibria from a unique coexistence equilibrium, to coexistence alongside a monomorphic \(A\)-equilibrium, and finally to a unique monomorphic \(A\)-equilibrium. Network structure alone therefore does not determine whether identities coexist: the relevant thresholds also depend on the behavioral environment induced by the population's characteristics. We next examine what determines these thresholds and how they respond to changes in peers' characteristics and the network structure.

\subsection{Identity thresholds and equilibrium persistence} \label{subsec:C_endogeneity}
Figure~\ref{fig:pure-equilibria} shows that changes in abilities can change equilibrium identity profiles even though social support at any given profile is unchanged. Such changes affect individuals' own valuations of identities and may also alter others' incentives through equilibrium peer behavior. We first state when a change in abilities or network links preserves an existing equilibrium, and then examine how peer behavior affects the relevant thresholds.

For a change in an exogenous variable, let \(\Delta\) denote the difference between post- and pre-intervention quantities, evaluated at the original identity profile. Each threshold uses the continuation equilibria under the two candidate identities, holding all other identities fixed; unmarked quantities denote initial values.

\begin{corollary}[Equilibrium persistence]
\label{cor:equilibrium-persistence}
Let \(s\) be a pure-strategy identity equilibrium. Following a feasible change in abilities or network links, with all other primitives fixed, \(s\) remains an equilibrium if and only if, for every individual \(i\), writing \(I=s_i\), and every \(J\neq I\),
\begin{equation}
\label{eq:equilibrium-persistence}
\Delta C_i(I,J,s_{-i})-\Delta(d_{i,I}-d_{i,J})
\leq d_{i,I}-d_{i,J}-C_i(I,J,s_{-i}).
\end{equation}
\end{corollary}

The right-hand side of equation~\eqref{eq:equilibrium-persistence} is individual~\(i\)'s initial payoff advantage from choosing \(I\) rather than \(J\), divided by \(\beta\). The left-hand side measures the reduction in that advantage following the intervention. The
original profile remains an equilibrium precisely when no such reduction exceeds the corresponding initial advantage. An intervention can therefore disrupt an equilibrium even when it leaves direct social support unchanged, by changing the relative core values of identities. We now elaborate on what exactly affects this threshold.

\begin{proposition} [Pairwise switching costs] \label{prop:Cdeterminants}
    For every individual $i$, identity $I$, alternative identity $J \neq I$ and strategy profile $s_{-i}$, the pairwise switching cost $C_i(I,J,s_{-i})$ satisfies the following:
    \begin{enumerate}[(i)]
        \item It is strictly decreasing in status $\mu_I$.
        \item It has a unique minimum in centrality $b_{i,I}$,
        \[ 
        \arg\min_{b_{i,I}} C_i(I,J,s_{-i}) = \frac{w_i}{1+\gamma w_i}. 
        \]
        
        \item It has a unique minimum in prescribed norm $v_I$,
        \[
        \arg\min_{v_I\geq 0} C_i(I,J,s_{-i}) = \max \{0, \frac{\frac{w_i}{1+\gamma w_i}-\frac{1-\alpha\psi_i}{\alpha\psi_i^2} \left(b_{i,I} - \frac{w_i}{1+\gamma w_i} \right)^2}{1-\gamma \left[\frac{w_i}{1+\gamma w_i}-\frac{1-\alpha\psi_i}{\alpha\psi_i^2} \left(b_{i,I} - \frac{w_i}{1+\gamma w_i} \right)^2\right]}\}.
        \]
    \end{enumerate}    
\end{proposition}

Part~\emph{(i)} captures the direct benefit of higher status. Part~\emph{(ii)} characterizes behavioral fit: an identity's core value is highest when the individual's centrality equals their peer-free benchmark, \(w_i/(1+\gamma w_i)\). Since equilibrium
actions are proportional to centrality, this is also the point at which their action equals their peer-free action. Part~\emph{(iii)} characterizes the prescribed norm that maximizes overall core value. Because the norm affects both the conformity target and equilibrium actions, this optimum need not minimize the norm-conformity loss itself. Corollary~\ref{cor:peer-ability}
connects behavioral fit to changes in another individual's ability. Let
\[
x_i^{\mathrm{pf}}(I)
\equiv
(1+\gamma v_I)\frac{w_i}{1+\gamma w_i}
\]
denote \(i\)'s peer-free action under identity \(I\).   

\begin{corollary}[Peer ability and identity attractiveness]
\label{cor:peer-ability}
\mbox{}\\
Fix the network $\mathbf{g}$, identity profile \(s_{-i}\), and all primitives except \(w_j\), where \(j\neq i\) holds identity \(I\) and belongs to the same connected within-\(I\) component as \(i\) when \(i\) chooses \(I\).

An increase in \(w_j\) strictly raises \(x_i^*(I,s_{-i})\). For every \(J\neq I\), it decreases \(C_i(I,J,s_{-i})\) if it moves \(x_i^*(I,s_{-i})\) closer to \(x_i^{\mathrm{pf}}(I)\), and increases \(C_i(I,J,s_{-i})\) if it moves \(x_i^*(I,s_{-i})\) farther away. The pairwise cost is unchanged if the distance from \(x_i^{\mathrm{pf}}(I)\) is unchanged.
\end{corollary}

In essence, an increase in another individual's ability\footnote{Assuming that both individuals belong to the same connected component of the network induced by identity \(I\).} raises individual \(i\)'s equilibrium action when identities are held fixed. These are standard peer effects. However, this increase also impacts the relative attractiveness of identity $I$. In particular, if it moves individual~$i$'s action under \(I\) farther from their peer-free action, \emph{ceteris paribus}, it makes it more costly for $i$ to choose $I$, thus increasing $\Delta C_i(I,J,s_{-i})$. If \(i\) initially chooses \(I\),
Corollary~\ref{cor:equilibrium-persistence} implies that the original profile ceases to be an equilibrium when the increase in \(C_i(I,J,s_{-i})\) for some \(J\neq I\) exceeds their initial payoff advantage from \(I\) over \(J\), divided by
\(\beta\). The resulting identity and action choices depend on how others subsequently adjust. It is possible that $i$'s action level decreases. 
We illustrate the idea that changes in underlying characteristics can disrupt an identity profile and lead to equilibrium behavior opposite to the one predicted by either fixed identities or fixed identity thresholds with the following examples.

First, note that changes in network structure can generate the same forces on actions and identity choices as changes in abilities. \citet{ushchev_zenou} identify conditions under which adding a link in a network increases every individual's equilibrium action in a local-average model. The following example considers a simple network of three individuals and adds exactly such a link, i.e., one that raises all equilibrium actions under fixed identities. It shows how endogenous identity choice reverses this effect on one individual's action.

\begin{example}[A link-induced equilibrium action reversal]
\label{ex:link-reversal}
\mbox{}\\
There are three individuals with abilities \((w_1,w_2,w_3)=(1,3,3)\) and two available identities, \(\mathcal I=\{A,B\}\). Table \ref{tab:link-reversal} compares the optimal action and identity profiles under three scenarios: First, the network is a line on which individual~1 is linked to individuals~2 and~3, who are not linked to each other. Second, the link \(\{2,3\}\) is added, but identities are kept fixed at the initial equilibrium and only actions are revised. Finally, both actions and identity choices are allowed to adjust in response to the added link.

\begin{table}
\centering
\small
\begin{threeparttable}
\caption{Network structure and equilibria} \label{tab:link-reversal}
\begin{tabular}{@{}lccc@{}}
\hline
Comparison & Identities & \(x_1^*\) & \(x_2^*=x_3^*\) \\
\hline
Original network equilibrium & \((A,A,A)\) & 1.634 & 1.859 \\
Added link, identities held fixed & \((A,A,A)\) & 1.736 & 2.038 \\
Added link, equilibrium identities & \((B,A,A)\) & 0.833 & 2.400 \\
\hline
\end{tabular}
\begin{tablenotes}
    \footnotesize
    \raggedright
    \item[] \textit{Notes:} Identity profiles and equilibrium actions under line and closed triangle network with parameter values $\alpha=2$, $\beta=0.1$, $\gamma=0.5$, $v_A=2$, $v_B=0.5$, $\mu_A=0.1$, $\mu_B=0$. 
    \end{tablenotes}
\end{threeparttable}
\end{table}

The original and final identity equilibria are unique. Although the link raises every action conditional on the original identities, it makes individual~1 prefer identity \(B\). Individual~1's equilibrium action consequently falls below its original level. Appendix~\ref{app:link-example} verifies the calculations.
\end{example}

Individual~1 illustrates the mechanism described in
Corollary~\ref{cor:peer-ability}: their equilibrium action under \(A\) exceeds their peer-free action. The added link moves individual~1's conditional action further above this benchmark, lowering the core value of \(A\) and increasing $C_i(I,J,s_{-i})$. This change induces individual~1 to change identity, despite no change in social support. Note that the possibility of changing identities in itself, with a fixed threshold, would leave individual~1's identity choice unchanged: the social support for individual~1 has not changed, so fixed thresholds imply that equation \eqref{eq:equilibrium-persistence} is trivially satisfied.\footnote{The action reversal is individual: aggregate action rises in this example. This is not generically the case. Within the same example, setting $\gamma=1$ with all other parameters unchanged causes a decrease in aggregate actions.}

\subsection{Social support and identity adoption}
\label{sec:composition-adoption}
The preceding analysis shows how changes in peers' characteristics or connections can alter an individual's identity choice even when their direct social support remains unchanged. The result of Corollary \ref{cor:equilibrium-persistence} is however more general. It specifically accommodates cases in which both social support and identity thresholds change. We now consider such a scenario. An additional adopter or a rewiring of connections across identities changes both how many neighbors of an individual share an identity with them and the equilibrium behavior associated with choosing it. 

With fixed identity thresholds, an additional adopter weakly increases others' incentives to choose the same identity. Likewise, a rewiring that replaces cross-identity links with same-identity links strengthens support for current identities and preserves an existing equilibrium. The following examples show how endogenous thresholds can overturn these predictions. An additional adopter can discourage another individual from choosing that identity, while greater homophily in links can make leaving one's current identity optimal. In both cases, the change in the behavioral consequences of identity choice outweighs the change in direct social support, thus causing the condition in Corollary \ref{cor:equilibrium-persistence} to be violated and the original identity profile to be disrupted.

\begin{example}[{\small An additional adopter can reduce equilibrium adoption}]
\label{ex:nonmonotone-adoption}
\mbox{}\\
Consider a complete network of four individuals, each with a unique ability level, and two identities, \(\mathcal I=\{A,B\}\), with an equilibrium identity profile \((A,A,A,B)\) as depicted in panel $(a)$ of Figure \ref{fig:composition-adopter}. Now consider an exogenous intervention that fixes individual~4's identity at \(A\) and allows the others to adjust their behavior, depicted in panel $(b)$. 

Rather than reinforcing adoption of $A$, this change incentivizes individual~1 to change their identity to $B$, which in turn leads to 2 also adjusting their identity. At this point, individual 4 endogenously prefers to retain identity \(A\). The resulting equilibrium identity profile is \((B,B,A,A)\), as depicted in panel $(c)$. Appendix~\ref{app:adoption-example} provides the underlying calculations.

\begin{figure}[!htbp]
 \centering
 \includegraphics[width=\linewidth]{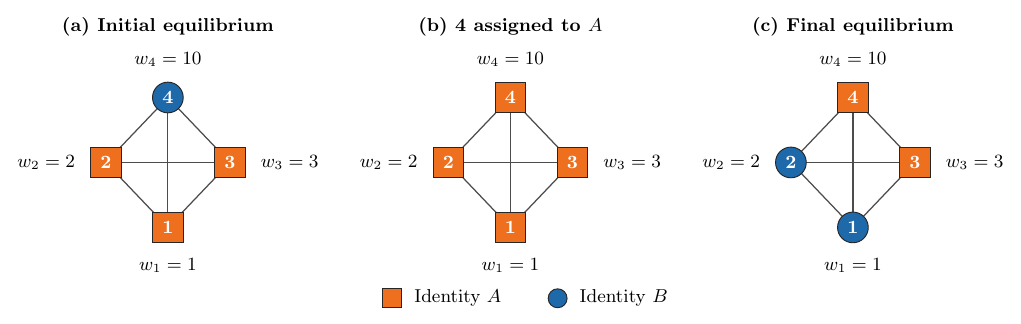}
 \caption{\textbf{An additional adopter reduces equilibrium adoption.}}
 \label{fig:composition-adopter}
 \figurenote{Orange squares denote \(A\) and blue circles denote \(B\). Parameters are \((\alpha,\beta,\gamma)=(0.5,0.01,0.1)\), \((\mu_A,\mu_B)=(1.5,1.5)\), and \((v_A,v_B)=(1,0)\). Panels~(a) and~(c) are strict identity equilibria. Panel~(b) assigns individual~4 to \(A\); the successive strict best responses of individuals~1 and~2 lead to panel~(c), where individual~4 prefers to retain \(A\).}
\end{figure}
\end{example}

Example \ref{ex:nonmonotone-adoption} highlights how an increased social support for an identity can be outweighed by changes in identities' core value. The mechanism is neighborhood composition. Individual~4's ability is significantly higher than that of individual~1, which implies that before the intervention they would have faced a significantly higher neighbor action if they had chosen identity $B$. After the intervention, this high action pressure has shifted to $A$, while $B$ now does not carry any peer pressure at all. This changes relative attractiveness of the two identities to an extent that it is beneficial for 1 to change their identity to $B$. Their own change then triggers individual~2 to prefer $B$ as well. With fixed identity thresholds, by contrast, the extra adopter would reinforce the choices of all three existing \(A\)-holders.\footnote{The initial equilibrium also illustrates that higher-ability individuals need not choose identities with higher prescribed norms. Individual~4 chooses \(B\), while all three lower-ability individuals choose \(A\). Joining \(A\) would expose 4 to conformity pressure from substantially lower-action peers, making it less attractive despite its higher prescribed norm.} 

Our next example also destabilizes an existing equilibrium, in this case through a rewiring that strengthens support for current identities.
\bigskip

\noindent \begin{example}[Homophily and endogenous identity change]
\label{ex:homophily-adoption}
\mbox{}\\
Consider the eight-individual network in panel~$(a)$ of Figure~\ref{fig:composition-homophily}. Individuals~1--4 have ability \(w_l=1\), individuals~6 and~8 have ability \(w_m=2\), and individuals~5 and~7 have ability \(w_h=10\). There are two identities, \(\mathcal I=\{A,B\}\). The identity profile depicted in panel~$(a)$ is a strict equilibrium: individuals~1--4 choose \(B\), while individuals~5--8 choose \(A\).

Panel~$(b)$ rewires the network by removing links \(\{1,5\}\) and \(\{3,7\}\) and adding links \(\{1,3\}\) and \(\{5,7\}\). The rewiring preserves every individual's degree, keeps the network connected, and increases homophily both by ability and by the initial identities. We assume that individuals update their identity myopically to the current identity profile. Both individuals~1 and~3 now strictly prefer \(A\), as depicted in panel~$(b)$. Their adoption subsequently makes \(A\) optimal for individuals~2 and~4 as well, yielding the all-\(A\) profile in panel~$(c)$, which is a strict equilibrium of the rewired network. Appendix~\ref{app:homophily-example} verifies the equilibrium comparisons and the resulting cascade.

\begin{figure}[htbp]
 \centering
 \includegraphics[width=0.9\linewidth]{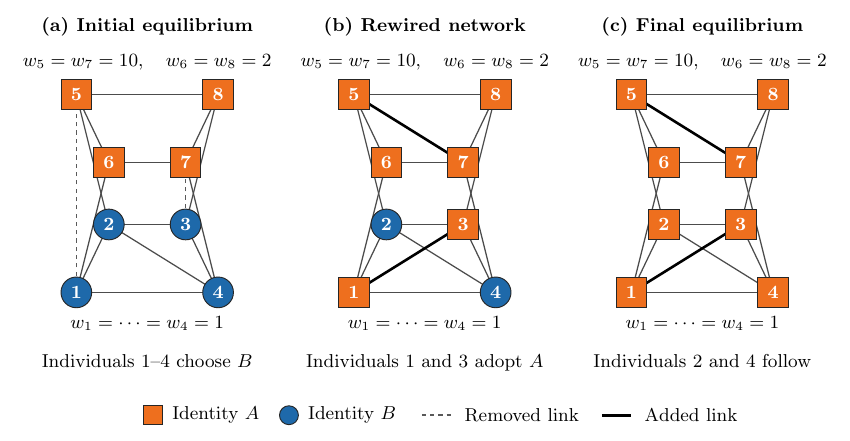}
 \caption{\textbf{Homophily and endogenous identity change.}}
 \label{fig:composition-homophily}
 \figurenote{Orange squares denote \(A\) and blue circles denote \(B\). Individuals~1--4 have ability \(1\), individuals~6 and~8 ability \(2\), and individuals~5 and~7 ability \(10\). The rewiring removes \(\{1,5\}\) and \(\{3,7\}\), shown dashed in panel~(a), and adds \(\{1,3\}\) and \(\{5,7\}\), shown thick in panels~(b) and~(c). Panel~(a) shows the initial strict equilibrium, panel~(b) the first simultaneous response to the rewiring, and panel~(c) the final equilibrium. Parameters are \((\alpha,\beta,\gamma)=(0.5,0.05,0.3)\), \((\mu_A,\mu_B)=(2.4,0)\), and \((v_A,v_B)=(5,0)\).}
\end{figure}
\end{example}

The cascade in Example \ref{ex:homophily-adoption} operates through changes in the composition, rather than the disappearance, of behavioral reference groups. For individuals~1 and~3, the rewiring removes a very high-ability \(A\)-neighbor while leaving an intermediate-ability \(A\)-neighbor in place. Their direct social support shifts toward \(B\), but the behavior they would face after adopting \(A\) becomes sufficiently less costly that both switch. If their pre-rewiring identity thresholds were held fixed, neither would adopt. Once they choose \(A\), individuals~2 and~4 still retain a \(B\)-neighbor, but the newly adopted low-ability members moderate equilibrium behavior in the prospective \(A\)-group. This change in core values, together with the increased direct support for \(A\), induces the second wave. Holding the thresholds of individuals~2 and~4 fixed at their pre-cascade values would again prevent adoption. Thus, both stages of the cascade require the endogenous behavioral response of identity values, even though every pivotal comparison involves nonempty reference groups on both sides.\footnote{In Examples~\ref{ex:link-reversal} and~\ref{ex:nonmonotone-adoption}, pivotal individuals instead evaluate switching to an identity without same-identity neighbors. This is not a necessary assumption for the illustrated mechanisms to work.}

Examples~\ref{ex:link-reversal}--\ref{ex:homophily-adoption} establish that endogenous identity thresholds can overturn predictions based on fixed identities or fixed adoption thresholds. The analytical results characterize when a given identity profile is disrupted; the examples establish the resulting reversals in the specified settings. Their implications depend on the ability distribution as well as the network. To isolate network effects operating through direct social support, we now impose additional structure.

\section{Homogeneous Society} \label{sec:homogeneity}
\subsection{Coexistence of identities and density}
\label{subsec:homogeneous}
The preceding section has examined how differences in abilities interact with network structure to shape identity choices. We now consider a population with identical abilities to study
how/whether network structure can sustain differences in identities and behavior among otherwise identical individuals. Formally, we impose the following assumption.

\begin{assumption}[Homogeneity] \label{assume:homogeneity}
    The entire population is homogeneous in abilities, such that $w_i=w$ for all individuals $i$.
\end{assumption}

Homogeneity yields the following result regarding equilibrium action levels and centralities.

\begin{lemma}[Actions and core values under homogeneity] \label{lem:homo_actions}
\mbox{}\\
Under Assumption~\ref{assume:homogeneity}, centralities and equilibrium actions in any identity $I$ satisfy
\[
b_{i,I}= \frac{w}{1+\gamma w} \qquad \Rightarrow \qquad 
x_{i,I}^*=[1+\gamma v_I]\frac{w}{1+\gamma w}\quad\text{for all }i.
\]
Consequently, the pairwise cost $C_i(I,J,s_{-i})$ is identical across individuals and independent of the identity profile $s_{-i}$; we denote this common value by $C(I,J)$.
\end{lemma}

Identical centralities under homogeneous abilities arise as individuals compare their chosen action to the \textit{average} action of their neighbors.\footnote{This result also arises in \cite{ushchev_zenou} and is in contrast with models such as \cite{BCZ}.} This directly implies identical actions within each identity, so peer-conformity losses vanish and the core values of identities are ranked identically by all. This directly implies the following corollary.

\begin{corollary}[Existence of pure-strategy equilibria] \label{coroll:homogeneous_equilibrium}
\mbox{}\\
Under homogeneity, there exists at least one identity $I^*$ with a maximal core value. The all-\(I^*\) profile is a pure-strategy equilibrium and maximizes aggregate equilibrium welfare.
\end{corollary}

It follows immediately that under identical abilities, any equilibria in which identities with different core values coexist must be sustained by differences in local identity supports, and that these equilibria are Pareto dominated by one in which all choose $I^*$: each individual obtains the maximal core utility and simultaneously maximizes the number of same-identity neighbors. To focus on how a network might sustain a coexistence equilibrium with different core value identities, we focus on two identities. 

\begin{assumption}[Two Identities] \label{assume:two_identities}
There are two identities, $\mathcal I=\{A,B\}$ with $A$ having the higher core value, such that $C(A,B)< 0$.
\end{assumption}

The all-\(A\) profile is therefore always an equilibrium. To characterize the persistence of \(B\), we measure each individual's support from a group relative to its complement.

\begin{definition}[Degree difference]
\label{def:k(S)}
For \(S\subseteq N\), write \(S^C=N\setminus S\) and define
\[
    k_i(S)=d_i(S)-d_i(S^C)
          =d_i\bigl[2\theta_i(S)-1\bigr],
    \qquad
    \theta_i(S)=\frac{d_i(S)}{d_i}.
\]
\end{definition}

The degree difference measures relative embeddedness: it is positive when an individual has more neighbors inside \(S\) than outside it.\footnote{Related degree-difference conditions appear in the literature on defensive \(k\)-alliances; see \citet{RodriguezVelazquezYeroSigarreta2009}. Here the condition measures the social support required to offset an identity's core-value disadvantage.} With \(S=N_B\), it is exactly the difference in social support entering the best-response condition of Lemma~\ref{lem:threshold}.

\begin{proposition}[Persistence of an inferior identity] \label{prop:homogeneous_coexistence}
\mbox{}\\
Under Assumptions \ref{assume:homogeneity} and 
\ref{assume:two_identities}:
    \begin{enumerate}[label=(\roman*)]
        \item The all-\(B\) profile is an equilibrium if and only if
            \[
                \min_{i\in N} d_i \ge -C(A,B).
            \]

        \item A coexistence equilibrium exists if and only if there exists a subset \(\varnothing \neq S \subsetneq N\) such that
            \[
                \min_{i\in S} k_i(S)\ge -C(A,B)
            \]
        and
            \[
                \min_{i\in S^C} k_i(S^C)\ge C(A,B).
            \]
        In this equilibrium, \(N_B=S\) and \(N_A=S^C\).
\end{enumerate}
\end{proposition}

Proposition \ref{prop:homogeneous_coexistence} specializes Theorem~\ref{prop:equilibria_pure_strategy} to a common core-value ranking. Every individual retaining \(B\) must have enough additional \(B\)-neighbors to offset its core-value disadvantage. Individuals choosing \(A\), by contrast, may do so despite a local majority of \(B\)-neighbors. Coexistence is therefore sustained by local social support, not differences in ability or core-value rankings. When \(v_A\neq v_B\), it also sustains different actions among otherwise identical individuals.

\begin{figure}
    \centering
    \includegraphics[width=0.8\linewidth]{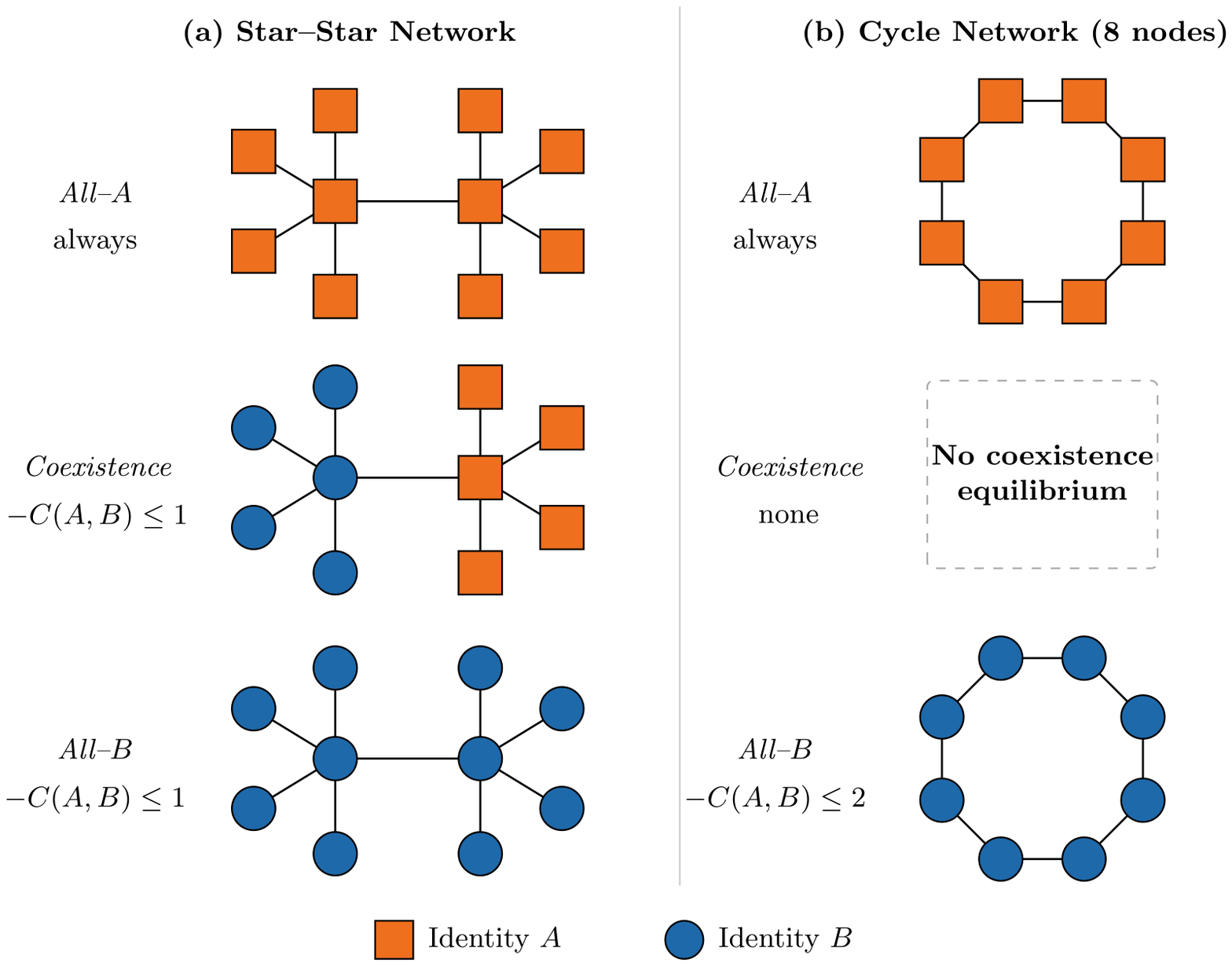}
    \caption{\textbf{Examples of equilibrium identity structures.}}
    \label{fig:mult_identity}
    \figurenote{Orange squares denote \(A\), which has the higher core value, and blue circles denote \(B\). Both networks admit an all-\(A\) identity profile. In panel~(a), \(\min_i d_i=1\), and both coexistence and an all-\(B\) profile are equilibria when \(0<-C(A,B)\leq1\). In panel~(b), \(\min_i d_i=2\), so all-\(B\) is an equilibrium profile when \(0<-C(A,B)\leq2\), while no coexistence equilibrium profile exists for \(C(A,B)<0\).}
\end{figure}

Figure~\ref{fig:mult_identity} distinguishes the persistence of \(B\) from coexistence. The regular cycle network in panel $(b)$ can sustain an all-\(B\) equilibrium over a wider range of core-value disadvantages than the two-stars network in $(a)$, yet it cannot sustain coexistence when \(A\) has the higher core value.

Here, a larger minimum degree enlarges the range of core-value disadvantages consistent with all-\(B\), but does not by itself ensure coexistence. More generally, degree also determines the strength of a given neighborhood composition: holding identity shares among neighbors fixed, a proportional increase in connections magnifies the difference between the numbers of neighbors supporting the two identities. The following corollary summarizes these two implications of network connectivity.

\begin{corollary}[Network structure and inferior identity persistence] \label{coroll:density}
\mbox{}\\
   Under Assumptions \ref{assume:homogeneity} and
\ref{assume:two_identities}:

\begin{enumerate}[(i)]
    \item As the minimum degree of a network increases, so does the range of core-value disadvantages for which an all-$B$ equilibrium exists. 

    \item Let a pure-strategy equilibrium exist in which both identities survive, with $S=N_B$ and $S^C=N_A$. Suppose the network undergoes a feasible proportional densification by a factor \(\lambda>1\) that leaves \(\theta_i(S)\) unchanged for every \(i\). Then there exists a pure-strategy equilibrium with \(N_B'\supseteq N_B\).
\end{enumerate}
\end{corollary}

Corollary~\ref{coroll:density} highlights two related roles of connectivity in sustaining the inferior identity. First, when everyone initially chooses \(B\), more connections make abandoning it more costly: the individual with the fewest neighbors determines how large a core-value disadvantage \(B\) can withstand. Second, when identities coexist, it is not degree alone but its interaction with neighborhood composition that matters. Proportional densification strengthens individuals' attachment to whichever identity is locally dominant, and therefore reinforces all individuals already choosing \(B\).

The asymmetry comes from \(A\)'s core-value advantage. An individual may optimally choose \(A\) despite having more \(B\)- than \(A\)-neighbors, whereas choosing \(B\) requires sufficiently stronger local support for \(B\). Densification can therefore destabilize some \(A\)-choices while it cannot destabilize existing \(B\)-choices through this channel. The corollary does not imply that every equilibrium on the denser network contains more \(B\)-adopters; rather, it shows that stronger social embeddedness can preserve, and potentially expand, an otherwise inferior identity. More generally, stronger embeddedness in a set makes identity choices less responsive to differences in core values. The same forces that sustain an inferior identity in equilibrium therefore also determine how easily such an equilibrium can be destabilized when relative identity attractiveness changes. We turn to this question next.

\vspace{0.75em}
\subsection{Diffusion of Identities} \label{subsec:diffusion}

The possibility that an inferior identity persists is economically consequential. Under homogeneity, the all-$A$ configuration Pareto dominates any configuration in which some individuals choose $B$, while differences in identity may generate persistent differences in actions even among individuals with identical abilities. We therefore ask how an increase in the relative attractiveness of $A$ can induce individuals to abandon $B$, and whether any initial changes of a subset of nodes propagate through the network. We impose a simple updating rule.

\begin{assumption}[Myopic updating]  \label{ass:myopic}
At each discrete step, all individuals simultaneously best respond to the current identity distribution in the network.
\end{assumption}

The best-response condition for \(A\) can be written as
\begin{equation}
\label{eq:threshold-degree}
    \theta_i(N_A)\geq\frac12+\frac{C(A,B)}{2d_i}.
\end{equation}
Since \(C(A,B)<0\), the required share of \(A\)-neighbors to induce a change from $B$ to $A$ is increasing in degree. Conditional on the same neighborhood identity shares, lower-degree individuals give greater weight to the core-value advantage of \(A\), whereas higher-degree individuals follow their local majority identity more closely. This degree dependence arises because core values do not scale with degree, while same-identity benefits accrue per connection.

This threshold characterizes an individual's choice for a given neighborhood composition. Once an individual changes identity, however, they also change the social support faced by their neighbors. An initial response to a change in relative identity attractiveness may therefore be amplified through the network. To study this process, we continue to assume that individuals update myopically in response to their current environment.

Suppose that the network initially rests in an all-$B$ equilibrium at some $C(A,B)< 0$. A common shock raises \(A\)'s core value, holding \(\widetilde V_B\) and \(\beta\) fixed, so that \(C'(A,B)<C(A,B)<0\). The network and post-shock primitives remain fixed during adjustment. To exclude indifference, assume that \(-C'(A,B)\) is unequal to every degree difference attainable in the network.\footnote{For a fixed finite network, this excludes only finitely many values of the post-shock pairwise cost.}

\begin{proposition}[Initiation and full diffusion] \label{prop:diffusion}
Under Assumptions \ref{assume:homogeneity}--\ref{ass:myopic}, consider an all-\(B\) equilibrium followed by a decrease in $C(A,B)$ to $C'(A,B)<C(A,B)$ as described. Adoption of \(A\) is irreversible along the adjustment path, and updating reaches a pure-strategy equilibrium after finitely many rounds. Moreover:

\begin{enumerate}[(i)]
\item Individual \(i\) adopts \(A\) in the first round if and only if
\[
    d_i<-C'(A,B).
\]
Consequently, diffusion begins if and only if
\[
    \min_{i\in N} d_i<-C'(A,B).
\]

\item Full diffusion occurs if and only if there exists no nonempty subset \(S\subseteq N\) such that
    \[
        k_i(S)\geq -C'(A,B) \qquad \text{for every } i\in S.
    \]

\end{enumerate}
\end{proposition}

Starting from all-\(B\), an individual switching alone sacrifices every same-identity connection. As \(A\)'s core-value advantage increases, the lowest-degree individuals are therefore the first to become willing to adopt. Once adoption starts, each switch reduces the support for \(B\) among the adopter's neighbors. Diffusion stops only when the remaining \(B\)-holders can resist the new core-value advantage even with everyone outside their group choosing \(A\).

Degree thus determines the initial response, while relative embeddedness determines whether this response develops into a wider cascade. The proposition does not rank individuals' influence or their adoption times after the first round. Appendix \ref{A:algorithm} provides a constructive algorithm to obtain the terminal identity configuration.\footnote{The procedure follows the iterative logic of core constructions such as that in \citet{gagnon-goyal}, with a degree-difference retention condition here rather than a common absolute within-group degree cutoff.} Unlike the heterogeneous case in Section~\ref{sec:identity}, further adoption cannot make \(A\) less attractive through a change in reference-group behavior: core values remain fixed throughout adjustment.

\begin{figure}  [htbp]
    \centering
    \includegraphics[width=0.65\linewidth]{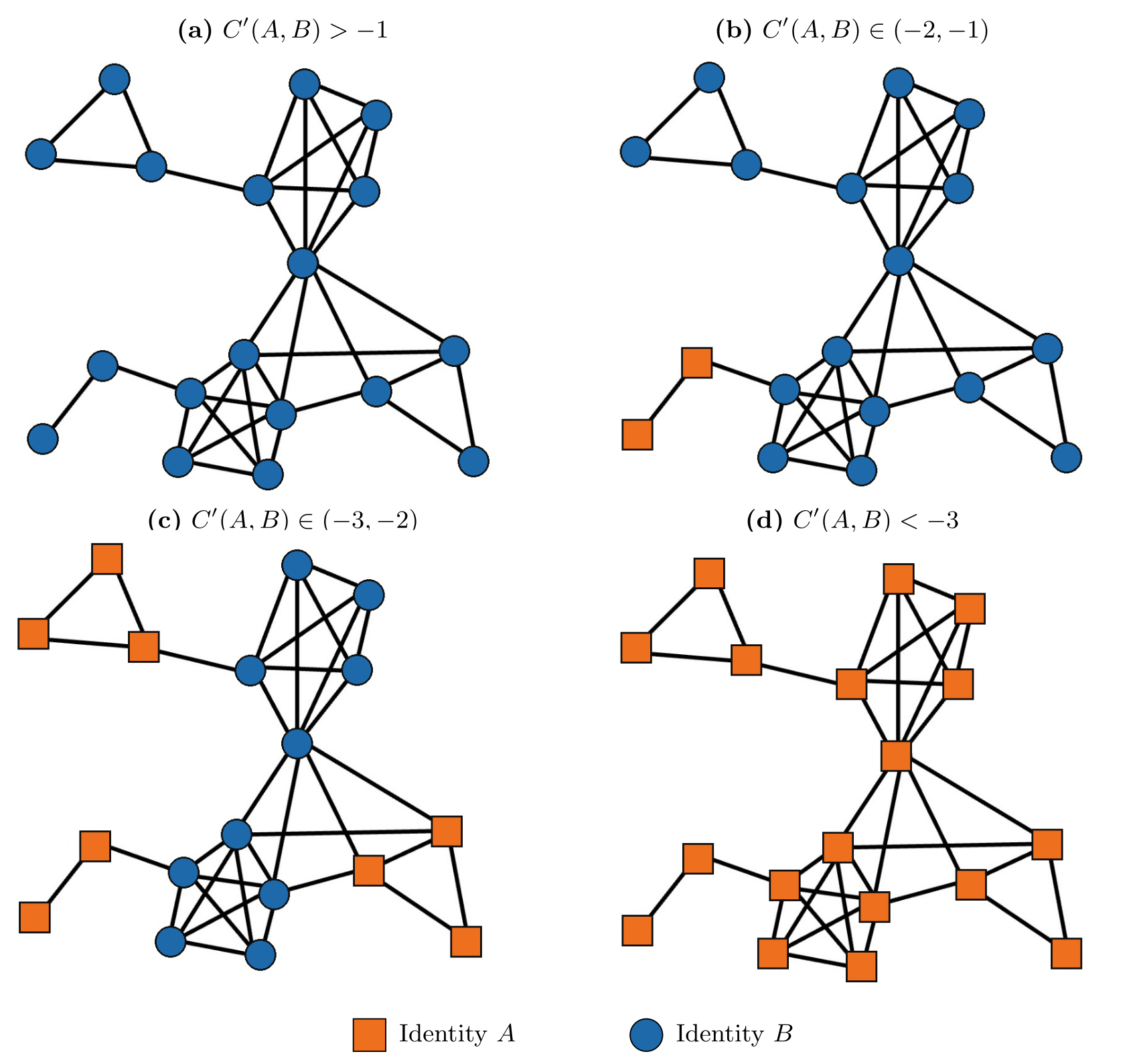}
    \caption{\textbf{Initiation and completion of identity diffusion.}}
    \label{fig:full_diffusion}
    \figurenote{Orange squares denote \(A\) and blue circles denote \(B\). Starting from the same all-\(B\) equilibrium, each panel reports the terminal identity configuration reached under myopic updating for the indicated range of \(C'(A,B)\). The panels correspond to distinct post-shock values, not consecutive updating rounds; threshold values at which an individual is indifferent are excluded, as in Proposition~\ref{prop:diffusion}.}
\end{figure}

Figure~\ref{fig:full_diffusion} illustrates the distinction between initiation and completion on a fixed network. Its minimum degree is one, so adoption begins only when \(-C'(A,B)>1\). Larger core-value advantages can induce further diffusion and reduce the group resisting adoption. The panels show terminal outcomes for different shock sizes, not successive rounds under one shock.

\subsection{Welfare effects} \label{subsec:welfare}
Although all-\(A\) is efficient, a transition toward it need not improve welfare. As before, consider a shock increasing the core value of $A$ and myopic adjustments. Since only \(A\)'s core value changes, the shock does not affect payoffs in the initial all-\(B\) profile before updating begins. Let \(W\) denote aggregate equilibrium utility in that original profile and \(W'\) its value in the terminal post-shock equilibrium. Write \(N_B'\) for the terminal \(B\)-group, \(N_A'=N\setminus N_B'\), and
\[
    \ell'=\sum_{i\in N_A'}d_i(N_B')
         =\sum_{i\in N_B'}d_i(N_A')
\]
for the number of links crossing the final identity partition.

\begin{proposition}[Welfare after identity diffusion]
\label{prop:welfare_diffusion}
\mbox{}\\
Under Assumptions \ref{assume:homogeneity}--\ref{ass:myopic} and a shock that increases the core value of $A$:
\begin{enumerate}[label=(\roman*)]
    \item The change in aggregate welfare is
    \begin{equation}
    \label{eq:homogeneous-welfare-change}
        W'-W=\beta\bigl[-C'(A,B)|N_A'|-2\ell'\bigr].
    \end{equation}
    Thus, welfare is strictly lower if and only if
    \begin{equation}
    \label{eq:welfare_loss}
        2\ell'>-C'(A,B)|N_A'|.
    \end{equation}
    \item Every individual retaining \(B\) is weakly worse off, with a strict loss precisely when they have a neighbor in \(N_A'\).
    \item An adopter \(i\in N_A'\) is strictly worse off if and only if
    \begin{equation}
    \label{eq:individual_welfare_loss}
        d_i(N_B')>-C'(A,B).
    \end{equation}
    Every first-round adopter is strictly better off in the terminal equilibrium than in the original all-\(B\) equilibrium.
\end{enumerate}
\end{proposition}

Each adopter gains the core-value advantage of \(A\), but each cross-identity link loses its same-identity benefit at both endpoints. Aggregate welfare falls when these losses outweigh the gains. A loss therefore requires partial diffusion: no adoption leaves welfare unchanged, while full adoption raises it. The condition stopping diffusion is different from the condition determining its welfare effect. The former concerns each remaining \(B\)-holder's degree difference; the latter concerns the total number of links crossing the terminal identity boundary.

The possibility of losses among adopters is not inconsistent with privately optimal updating. Later adopters respond to a social environment already changed by earlier switches. Adopting can be better than remaining with \(B\) in that environment while still yielding less utility than in the original equilibrium. The following example illustrates this.

\begin{example}[Welfare-reducing partial diffusion]
\label{ex:homogeneous-welfare}
\mbox{}\\
Consider the six-individual network in Figure~\ref{fig:social_mobility}. Let \(\beta=1\), normalize \(\widetilde V_B=0\), and raise the core-value advantage of \(A\) from \(-C(A,B)=0.2\) to \(-C'(A,B)=1.5\).

Initially, all-\(B\) is an equilibrium. Individual~4, with degree~1, adopts first. Individual~3 then has two \(B\)-neighbors and one \(A\)-neighbor, so their degree difference is \(1<1.5\) and they also adopt. Diffusion stops at \(A'=\{3,4\}\), since every member of \(B'=\{1,2,5,6\}\) has degree difference~2.

\begin{figure}[ht] 
    \centering
    \includegraphics[width=0.65\linewidth]{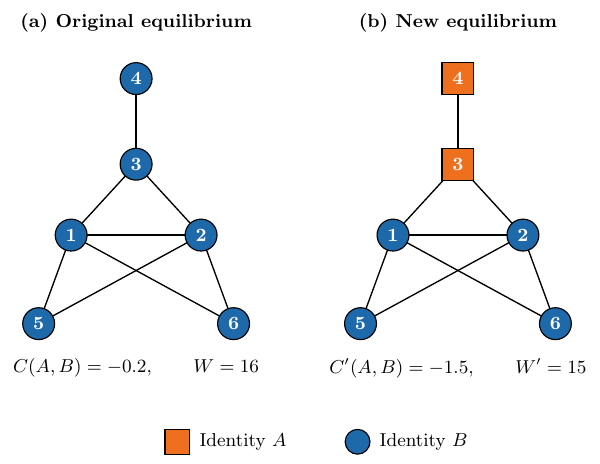}
    \caption{\textbf{Welfare-reducing partial diffusion.}}
    \label{fig:social_mobility}
    \figurenote{Orange squares denote \(A\) and blue circles denote \(B\). Panel~(a) shows the initial all-\(B\) equilibrium at \(C(A,B)=-0.2\); panel~(b) shows the terminal equilibrium after the shock to \(C'(A,B)=-1.5\), in which individuals~3 and~4 adopt \(A\) while the others retain \(B\). In both panels, \(\beta=1\) and \(\widetilde V_B=0\). The network is unchanged; the two links crossing the final identity partition remain present but no longer yield same-identity benefits.}
\end{figure}

Individual~3 obtains \(2.5\) after adopting, rather than the \(2\) they would obtain by retaining \(B\) after individual~4's switch. Both are below their original utility of~3. With \(|A'|=2\) and \(\ell'=2\), aggregate welfare falls from \(16\) to~\(15\).
\end{example}

The example separates the incentive to respond to others' identity changes from the welfare effect of the transition as a whole. A common improvement in one identity's core value can induce privately optimal changes that stop short of full coordination and leave society and changing individuals worse off than in the original equilibrium. These welfare comparisons do not themselves imply higher actions: by Lemma~\ref{lem:homo_actions}, adoption raises actions precisely when \(v_A>v_B\).

\section{Applications to peer effects} \label{sec:applications}
\subsection{Economic connectedness and educational identity}
\label{subsec:connectedness}

Our model distinguishes between exposure to a peer and that peer's relevance for behavior. Identity determines which existing connections enter an individual's action reference group, while those connections affect the attractiveness of different identities. We illustrate this mechanism through two applications: economic connectedness and the effects of deliberately designed peer groups. The first is developed in the present subsection and uses the homogeneous benchmark to isolate changes in educational identification; the second (Section~\ref{subsec:designed_peers}) shows how heterogeneous abilities affect both identity choice and behavior within a given identity.

\citet{chetty_nature1} document that economic connectedness---the prevalence of friendships between individuals with low and high socioeconomic status (SES)---is among the strongest known predictors of upward economic mobility. Our model offers a possible mechanism linking these friendships to educational investment. Here, the exogenous network represents existing friendships, and identity determines which friends constitute behavioral reference points. Socioeconomic background and educational identity are distinct: the former describes an individual's family background, whereas the latter is chosen.

Let \(\mathcal{I}=\{H,L\}\), where \(H\) and \(L\) denote educational identities with prescribed investment levels \(v_H>v_L\). Consider the homogeneous-ability benchmark and suppose that \(H\) also has the higher core value, so that \(C(H,L)<0\). Assume that students from high-SES backgrounds are initially more likely to hold identity \(H\), for example because of differences in parental expectations.\footnote{This initial association is consistent with evidence that disadvantaged students hold lower educational aspirations than equally achieving advantaged students \citep{guyon2021biased}. We take the association as given rather than model its formation; educational identity may subsequently change through social interaction.}

The adoption condition in equation~\eqref{eq:threshold-degree}, derived in Section~\ref{subsec:diffusion}, then applies directly: an individual strictly prefers \(H\) when the share of \(H\)-identifying friends exceeds a degree-dependent threshold. Consider a low-SES student initially choosing \(L\). Holding their degree and other individuals' identities fixed, replacing an \(L\)-identifying friend with an \(H\)-identifying high-SES friend raises the share of friends supporting \(H\). If this change crosses the threshold, the student adopts \(H\), increasing investment by
\begin{equation}
    x_{i,H}^{*}-x_{i,L}^{*}
    =\frac{\gamma w}{1+\gamma w}(v_H-v_L)>0.
    \label{eq:identity_investment}
\end{equation}
In this benchmark, the action increase comes from the higher prescribed norm: conditional on identity, equilibrium actions are independent of peer composition. The network matters because it determines whether that identity is adopted. If future earnings increase with educational investment, cross-class friendships can therefore improve earnings prospects without requiring differences in ability, information provision, material assistance, or job referrals.

This mechanism yields two implications. First, identity adoption and investment respond discontinuously when exposure crosses the threshold. Consequently, the effect of economic connectedness depends on how cross-class friendships are distributed: connections that move students across their thresholds matter more for their investment than connections to students who already choose \(H\) or remain below the threshold. A higher average level of connectedness alone does not establish that more students cross their thresholds. Second, because \(C(H,L)<0\), the adoption threshold is increasing in degree. Conditional on the same share of \(H\)-identifying friends, lower-degree students are more inclined to adopt \(H\), whereas higher-degree students follow their local majority more closely.

The model also helps explain why \citet{chetty_nature1} find that economic connectedness predicts upward mobility more strongly and consistently than structural measures of network cohesion. In our framework, cohesive neighborhoods can sustain low educational investment by stabilizing identity \(L\), just as they can sustain high investment by supporting \(H\). Structural cohesion alone therefore does not identify the behavioral implications of social support. Under the assumed association between socioeconomic background and educational identity, economic connectedness provides additional information: it captures connections to individuals more likely to support \(H\), whose adoption raises educational investment. The model thus identifies a mechanism through which strong local social ties can coexist with limited mobility, while cross-class friendships can facilitate upward mobility.

A complementary finding concerns peer effects beyond observed achievement. \citet{CornoLaFerraraBurns2022} find that interracial roommate assignment improves Black students' academic performance even after controlling for roommates' admission scores, while detecting no performance effect for White students. Our model identifies educational identity as an additional dimension of peer exposure: two equally able peers can have different effects because they support different identities. In the homogeneous benchmark, replacing an \(L\)-identifying peer with an \(H\)-identifying peer increases the social support for \(H\) without changing peer ability. If this replacement induces an individual to adopt \(H\), educational investment rises through the change in the prescribed norm. Thus, controlling for peer achievement need not account for the characteristics through which peers influence educational choices. Applying this mechanism to Black students' gains requires interracial assignment to increase their exposure to \(H\)-identifying peers conditional on prior achievement. Such differences in educational identification could reflect the influence of racial stereotypes; race itself remains distinct from the chosen identity. This interpretation yields a further empirical implication: conditional on roommate race and achievement, the effect of assignment should depend on the roommate's educational identification. Gains operating through this channel should accompany changes in the student's own educational identification and behavioral reference group.

\subsection{Designed peer groups and identity adjustment}
\label{subsec:designed_peers}

\citet{carrell2013} provide a useful setting for examining the distinction between assigned and effective peers. Drawing on positive peer effects estimated under random assignment, they designed groups to improve the academic performance of low-achieving students. The intervention placed these students with peers who had high SAT Verbal scores and assigned middle-achieving students to comparatively homogeneous groups. Performance declined among low-achieving students, improved among middle-achieving students, and did not change significantly among high-achieving students. Social and study groups also became more homogeneous.\footnote{The assignment algorithm distinguished predicted achievement from SAT Verbal scores. The example below uses a single ability dimension and illustrates the qualitative findings; its parameters are not estimated from these data, and its actions are not measured in GPA units.}

Our model illustrates how a change in peer composition can raise actions conditional on identities while inducing an identity change that reverses this gain. It shows that endogenous identity and reference-group choice provide one mechanism capable of generating the qualitative findings in \citet{carrell2013}. Consider sixteen students with abilities \((w_{\ell},w_m,w_h)=(1,5,5.25)\), initially assigned to two complete groups with compositions \((4,4,0)\) and \((0,4,4)\), where entries count low, middle, and high types. Students choose between identities \(\mathcal{I}=\{H,L\}\), with \(H\) providing higher status and prescribing a higher norm. Only peers sharing a student's identity enter their effective reference group. Under the parameters in Table~\ref{tab:carrell}, all students choose \(H\) in a strict equilibrium in each benchmark group.\footnote{Concretely, we apply our model to each assigned peer group separately.}

Table~\ref{tab:carrell} reports identity profiles and the associated action equilibria. Replacing one middle type with a high type in the first benchmark group raises low types' actions while preserving the all-\(H\) equilibrium. The designed assignment instead reallocates the original sixteen students into a bimodal group \((4,0,4)\) and a middle-only group \((0,8,0)\), preserving the population's ability composition. If each student's identity threshold is held at its benchmark value, all students retain \(H\), and low types' actions rise further. The fixed-\(C\) row reports this counterfactual; its all-\(H\) profile is not an identity equilibrium under endogenous thresholds.

\begin{table}[htbp]
    \centering
    \caption{Peer composition, identity adjustment, and equilibrium actions}
    \label{tab:carrell}
    \small
    \setlength{\tabcolsep}{4pt}
    \begin{tabular}{@{}lccrrr@{}}
        \hline
        Configuration & Ability & Identity
        & \(x_{\ell}\) & \(x_m\) & \(x_h\) \\
        & composition & profile & & & \\
        \hline
        Benchmark 1
        & \((4,4,0)\) & \((H,H,\text{--})\)
        & 1.620 & 3.197 & -- \\
        Benchmark 2
        & \((0,4,4)\) & \((\text{--},H,H)\)
        & -- & 5.068 & 5.127 \\
        Higher-\(h\) group
        & \((4,3,1)\) & \((H,H,H)\)
        & 1.623 & 3.204 & 3.242 \\
        Bimodal, fixed \(C\)
        & \((4,0,4)\) & \((H,\text{--},H)\)
        & 1.634 & -- & 3.263 \\
        Bimodal group
        & \((4,0,4)\) & \((L,\text{--},H)\)
        & 0.976 & -- & 5.198 \\
        Middle-only group
        & \((0,8,0)\) & \((\text{--},H,\text{--})\)
        & -- & 5.000 & -- \\
        \hline
    \end{tabular}
    \par\smallskip
    \begin{minipage}{\linewidth}
        \footnotesize
        \emph{Notes:} Ability composition reports
        \((n_{\ell},n_m,n_h)\); identity profiles report the
        corresponding types' identities. Parameters are
        \((w_{\ell},w_m,w_h)=(1,5,5.25)\),
        \((\alpha,\beta,\gamma)=(0.5,0.09,0.05)\),
        \((\mu_L,\mu_H)=(0,0.06)\), and
        \((v_L,v_H)=(0.5,5)\).
        The fixed-\(C\) row holds each student's threshold at its benchmark value.
    \end{minipage}
\end{table}

Allowing thresholds to adjust reverses the effect on low types. At the initial all-\(H\) profile in the bimodal group, replacing middle-type peers with high types raises the action associated with \(H\) and reduces its attractiveness relative to \(L\). Low types' threshold \(C_i(H,L,s_{-i})\) rises from \(6.832\) to \(7.053\), while their link difference \(d_{i,H}-d_{i,L}\) remains seven. Thus, \(L\) becomes their strict best response despite unchanged social support for \(H\). Under simultaneous best-response updating, all low types switch to \(L\) in one round, while middle and high types retain \(H\). The resulting profiles are strict equilibria; Appendix~\ref{A:carrell} verifies the comparisons.

Low types now face a lower prescribed norm and no upward peer pressure from high types, so their actions fall below the benchmark. High types gain because low types' departure removes downward peer pressure; their final action is only slightly above their original benchmark. Middle types remain in \(H\): those previously grouped with low types gain more than those previously grouped with high types lose, so their average action rises. Endogenous thresholds therefore generate both the reversal for low types and the segregation of effective peers in this calibration. The example illustrates a mechanism consistent with the qualitative findings; the evidence in \citet{carrell2013} does not identify these identity thresholds or their contribution to the treatment effect.

\section{Conclusions} \label{sec:conclusions}
This paper studies the interaction between endogenous identity formation and peer influence. Individuals are embedded in a fixed social network, but choose an identity that determines both the behavioral norm they face and which neighbors constitute their reference group. If individuals have heterogeneous abilities, the attractiveness of an identity depends crucially on the costs of conforming to its members' equilibrium behavior and individuals may choose to change their identity to reduce these costs. The consequences of this are significant. An intervention that raises actions within an existing identity may make that identity less attractive for some individuals, potentially causing identity changes and reduced actions instead. The same interaction between belonging and behavior explains why additional adopters of an identity can discourage future adoption, and why increasing homophilous ties can sometimes facilitate identity change.

A homogeneous-ability benchmark isolates the role of social attachment. Despite otherwise identical individuals, who rank identities' core values identically, network structure can sustain different identities and, through their different prescriptions, persistent differences in behavior. Yet a transition toward an identity with a higher core value need not improve welfare. If only some individuals switch, the loss of same-identity benefits on connections between adopters and non-adopters can outweigh the gains. Later adopters can even end up worse off than in the original equilibrium, although switching is optimal once others have changed identity.

The broader implication is that changes in social exposure can affect behavior along two margins: adjustment within an existing identity and identity changes that redefine the relevant norm and reference group. This distinction provides a common framework for interpreting empirical findings such as the effects of cross-group friendships, peer influence beyond observed ability, and the failure of estimated peer effects to predict the consequences of deliberate reassignment. For interventions based on peer exposure, network structure and observable peer characteristics (including abilities) may therefore be insufficient to predict outcomes. Information about identities, perceived norms, and reference groups may help explain which peers become influential and when changes in composition encourage the adoption of an alternative identity. Understanding social influence consequently requires attention to how individuals form the groups through which that influence operates.

\newpage
\appendix
\setcounter{equation}{0}
\renewcommand{\theequation}{A-\arabic{equation}}
\section{Proofs and numerical examples}

\subsection{Proof of Lemma \ref{lem:action_equ}} \label{A:existence_x*}
The derivation of equation \eqref{eq:equilibrium_x} follows standard arguments: Fix an identity $I$ and consider the best-response system implied by \eqref{eq:FOC} for all individuals with identity $I$:
\begin{equation}
    \mathbf{x_I} = [1+\gamma v_I]\mathbf{\psi_I} + \alpha\mathbf{\Tilde{G}_I}\mathbf{x_I},
\end{equation}
exploiting the fact that $\mathbf{\Tilde{G}_I}$ already includes $\psi_i$. Standard algebra leads to
\begin{equation} \label{Aeq:x}
    \left(\mathbf{I} - \alpha\mathbf{\Tilde{G}_I} \right)\mathbf{x_I} = [1+\gamma v_I]\mathbf{\psi_I}
\end{equation}
If a solution to equation \eqref{Aeq:x} exists, it is equation \eqref{eq:equilibrium_x}, and it is unique. Existence in turn is guaranteed by the fact that $\tilde{\mathbf G}_I$ is obtained from $\hat{\mathbf{G}}_I$, in which each non-zero row sums to one, by multiplying each row $i$ by a factor $\psi_i$, which implies that $\alpha\psi_i<1$ for every nonzero row. Therefore, denoting by $\rho(\tilde{\mathbf G}_I)$ the spectral radius of $\tilde{\mathbf G}_I$, it follows that $\alpha \rho(\tilde{\mathbf G}_I) < 1$, and that the matrix $(\mathbf{I} - \alpha \tilde{\mathbf G}_I)$ is invertible. 

Note that for a fixed identity profile, centralities are weakly increasing in abilities. Indeed, the spectral-radius condition implies
    \[
        b_I=\sum_{k=0}^{\infty} \alpha^k\widetilde G_I^{\,k}\psi_I.
    \]
An increase in \(w_j\) raises \(\psi_j\) and weakly increases
every entry of \(\widetilde G_I\), so every term in this nonnegative expansion weakly increases. For any individual \(i\) in the same connected within-identity component as \(j\), a path connecting them gives a strictly increasing contribution. Thus \(b_{i,I}\), and consequently \(x_{i,I}^*\), strictly increases. \hfill $\blacksquare$

\subsection{Proof of Lemma \ref{lem:threshold}}
\label{app:proof-threshold}
For any \(J\neq I\),
\[
\begin{aligned}
V_i(I,s_{-i})\ge V_i(J,s_{-i})
&\iff \beta(d_{i,I}-d_{i,J})
\ge \widetilde V_i(J,s_{-i})-\widetilde V_i(I,s_{-i})\\
&\iff d_{i,I}-d_{i,J}\ge C_i(I,J,s_{-i}),
\end{aligned}
\]
where the second equivalence follows from Definition~\ref{def:C_{i,I}} and \(\beta>0\). Identity \(I\) is a best response if and only if this comparison holds against every alternative \(J\neq I\). \hfill\(\blacksquare\)

\subsection{Proof of Theorem \ref{prop:equilibria_pure_strategy}}
\label{app:proof-pure-equilibria}
Let \(s\) induce the identity partition \(\{N_I\}_{I\in\mathcal I}\). For every \(i\in N_I\), Lemma~\ref{lem:threshold} implies that \(I\) is a best response given \(s_{-i}\) if and only if
\[
d_i(N_I)-d_i(N_J)\ge C_i(I,J,s_{-i})
\qquad\text{for every }J\neq I,
\]
where we used \(d_{i,I}=d_i(N_I)\) and \(d_{i,J}=d_i(N_J)\). Hence \(s\) is a pure-strategy Nash equilibrium if and only if these inequalities hold simultaneously for every individual in every nonempty identity group. If only one group \(N_I=N\) is nonempty, then \(d_i(N_I)=d_i\) and \(d_i(N_J)=0\) for every \(J\neq I\), yielding part~\emph{(i)}. If at least two groups are nonempty, the equilibrium exhibits coexistence, yielding part~\emph{(ii)}. \hfill\(\blacksquare\)

\subsection{Proof of Corollary \ref{cor:equilibrium-persistence}}
\label{app:proof-equilibrium-persistence}

Fix an individual \(i\), write \(I=s_i\), and consider any \(J\neq I\). By Lemma~\ref{lem:threshold}, \(I\) remains a best response after the intervention if and only if
\[
d_{i,I}-d_{i,J}+\Delta(d_{i,I}-d_{i,J})
\geq
C_i(I,J,s_{-i})+\Delta C_i(I,J,s_{-i})
\qquad\text{for every }J\neq I.
\]
Rearranging gives equation~\eqref{eq:equilibrium-persistence}. The original profile remains an equilibrium if and only if these inequalities hold for every individual and every alternative identity. \hfill\(\blacksquare\)

\subsection{Proof of Proposition \ref{prop:Cdeterminants}} 

Recall that $C_i(I,J,s_{-i})=1/\beta[\widetilde V_{i,J}-\widetilde V_{i,I}]$, i.e., for any $i$, $J\neq I$ and $s_{-i}$, the costs of switching from identity $J$ to $I$ are decreasing whenever $\widetilde V_{i,I}$ is increasing. Furthermore, recall that the core value of identity $I$ for individual $i$ is
\[
\widetilde V_{i,I} = \mu_I + x_{i,I}^* - \frac{(x_{i,I}^*)^2}{2w_i} -\frac{\alpha}{2}1\{d_{i,I}>0\}\left(x_{i,I}^*-\bar x_{i,I}\right)^2
-\frac{\gamma}{2}\left(x_{i,I}^*-v_I\right)^2
\]

Using the first order condition for $x_{i,I}$ and the equilibrium representation, we can write
\[
\bar x_{i,I} = \frac{1+\gamma v_I}{\alpha}\left(\frac{b_{i,I}}{\psi_i}-1\right).
\]

Substituting these equilibrium expressions into the utility function and simplifying yields
\[
\widetilde V_{i,I} = \mu_I + \frac12\left[(1+\gamma v_I)^2 \left\{\frac{b_{i,I}^2}{\psi_i} - \frac1\alpha \left(
\frac{b_{i,I}}{\psi_i} - 1\right)^2 \right\} - \gamma v_I^2
\right]
\]

and after further algebra
\begin{small}
\begin{equation}
    \widetilde{V}_{i,I} = 
        \mu_I + \frac{1}{2} \left[(1+\gamma v_I)^2 \frac{w_i}{1+\gamma w_i} - \gamma v_I^2 \right] -  \frac{(1+\gamma v_I)^2}{2}\frac{1-\alpha\psi_i}{\alpha\psi_i^2} \left(b_{i,I} - \frac{w_i}{1+\gamma w_i} \right)^2 
\end{equation}
\end{small}

Differentiating $\widetilde{V}_{i,I}$ with respect to $\mu_I$, $b_{i,I}$ and $v_I$ yields the results in Proposition \ref{prop:Cdeterminants}. Note that for $d_{i,I}=0$, centrality is automatically the peer-free (and optimum) centrality. Finally, note that it is possible for
\[
\frac{w_i}{1+\gamma w_i} - \frac{1-\alpha\psi_i}{\alpha\psi_i^2} \left(b_{i,I} - \frac{w_i}{1+\gamma w_i} \right)^2
\]
to be negative (in particular if an individual with low $w_i$ finds themselves in an identity that imposes a high centrality $b_{i,I}$). If this is the case, the switching costs are strictly increasing for all admissible values of $v_I\geq 0$, leading to the optimal $v_I$ being zero. \hfill $\blacksquare$

\subsection{Proof of Corollary \ref{cor:peer-ability}}
\label{app:proof-peer-ability}
As shown in the proof of Lemma~\ref{lem:action_equ}, an increase in \(w_j\) strictly raises \(b_{i,I}\), and hence
\[
x_i^*(I,s_{-i})=(1+\gamma v_I)b_{i,I},
\]
whenever \(i\) and \(j\) belong to the same connected within-\(I\) component. By Proposition~\ref{prop:Cdeterminants}, the core value of \(I\) is strictly decreasing in the distance
\[
\left|b_{i,I}-\frac{w_i}{1+\gamma w_i}\right|.
\]
Since \(1+\gamma v_I>0\), this distance increases, decreases, or remains unchanged if and only if
\[
\left|x_i^*(I,s_{-i})-x_i^{\mathrm{pf}}(I)\right|
\]
does likewise. The increase in \(w_j\) does not affect \(i\)'s core value under any \(J\neq I\). The result therefore follows from
Definition~\ref{def:C_{i,I}}. \hfill\(\blacksquare\)

\subsection{Verification of Example \ref{ex:link-reversal}}
\label{app:link-example}

The example uses \((w_1,w_2,w_3)=(1,3,3)\),\((\alpha,\beta,\gamma)=(2,0.1,0.5)\), \((\mu_A,\mu_B)=(0.1,0)\) and \((v_A,v_B)=(2,0.5)\). The original network \(\mathbf{g}\) contains links \(\{1,2\}\) and \(\{1,3\}\); \(\mathbf{g'}\) additionally contains \(\{2,3\}\).

Holding identities fixed at \((A,A,A)\), symmetry implies
\(x_2=x_3\equiv y\). On \(\mathbf{g}\), the action best responses reduce to
\[
x_1=\frac{4}{7}(1+y),
\qquad
y=\frac{12}{17}(1+x_1),
\]
yielding
\[
x^*(A,A,A;g)
=
\left(
\frac{116}{71},
\frac{132}{71},
\frac{132}{71}
\right)
\approx(1.634,1.859,1.859).
\]
After adding \(\{2,3\}\),
\[
x_1=\frac{4}{7}(1+y),
\qquad
y=\frac{6}{17}(2+x_1+y),
\]
and therefore
\[
x^*(A,A,A;g')
=
\left(
\frac{92}{53},
\frac{108}{53},
\frac{108}{53}
\right)
\approx(1.736,2.038,2.038).
\]
Thus the added link raises every action conditional on identities.

If individuals~2 and~3 choose \(A\), individual~1 has
\(d_{1,A}-d_{1,B}=2\) on both networks. Lemma~\ref{lem:threshold}
therefore implies that \(A\) is optimal if and only if
\(2\geq C_1(A,B,(A,A))\). The corresponding thresholds are
\[
C_1(A,B,(A,A);g)\approx1.435,
\qquad
C_1(A,B,(A,A);g')\approx2.376.
\]
Hence individual~1 strictly prefers \(A\) on \(\mathbf{g}\) but \(B\) on \(\mathbf{g}'\), despite unchanged direct social support.

For individuals \(i\in\{2,3\}\), evaluating all four possible profiles
of the other two identities gives
\[
\min_{s_{-i}}
\bigl[V_i(A,s_{-i};g)-V_i(B,s_{-i};g)\bigr]
\approx0.552>0,
\]
and
\[
\min_{s_{-i}}
\bigl[V_i(A,s_{-i};g')-V_i(B,s_{-i};g')\bigr]
\approx0.452>0.
\]
Thus both strictly prefer \(A\) on either network, so the unique identity equilibria are \((A,A,A)\) on \(\mathbf{g}\) and \((B,A,A)\) on \(\mathbf{g'}\).

At the latter profile,
\[
x^*(B,A,A;g')
=
\left(
\frac{5}{6},
\frac{12}{5},
\frac{12}{5}
\right)
\approx(0.833,2.400,2.400),
\]
which verifies the action reversal reported in Table~\ref{tab:link-reversal}. \hfill$\blacksquare$

\subsection{Verification of Example~\ref{ex:nonmonotone-adoption}}
\label{app:adoption-example}

The network is complete and the example uses \(
(w_1,w_2,w_3,w_4)=(1,2,3,10)\), $(\alpha,\beta,\gamma)=(0.5,0.01,0.1)$, \((\mu_A,\mu_B)=(1.5,1.5)\) and \((v_A,v_B)=(1,0)\).
All threshold values below are computed from the corresponding continuation equilibrium and rounded to three decimal places.

At the initial profile \((A,A,A,B)\), individuals~1--3 have
\(d_{i,A}-d_{i,B}=1\), while individual~4 has
\(d_{4,A}-d_{4,B}=3\). Their thresholds are
\[
\bigl(
C_1,C_2,C_3,C_4
\bigr)
=
(-30.688,-39.040,-24.842,43.540),
\]
so \((A,A,A,B)\) is a strict identity equilibrium.

Now assign individual~4 to \(A\), while all other individuals simultaneously best respond to the current identity profile each period. At \((A,A,A,A)\),
\[
d_{1,A}-d_{1,B}=3
    < C_1(A,B,(A,A,A))
    \simeq 30.642,
\]
so individual~1 strictly prefers \(B\). At the resulting profile
\((B,A,A,A)\),
\[
d_{2,A}-d_{2,B}=1
    < C_2(A,B,(B,A,A))
    \simeq 3.205,
\]
so individual~2 also switches to \(B\). At the resulting profile \((B,B,A,A)\), the relative social supports are
\[
(1,1,-1,-1),
\]
while the corresponding thresholds are
\[
(39.588,3.205,-7.682,-97.673).
\]
Hence individuals~1 and~2 strictly prefer \(B\), while individuals~3 and~4 strictly prefer \(A\), verifying the final equilibrium.

Finally, with individual~4 fixed at \(A\),
\[
\min_{s_2,s_3}
C_1\bigl(A,B,(s_2,s_3,A)\bigr)>17>3,
\]
so individual~1 strictly prefers \(B\) for every configuration of individuals~2 and~3. Conditional on \(s_1=B\),
\[
\min_{s_3}
C_2\bigl(A,B,(B,s_3,A)\bigr)>3.2>1,
\]
so individual~2 also strictly prefers \(B\). Individual~3 then strictly prefers \(A\). Thus \((B,B,A,A)\) is the unique equilibrium under the assignment. \hfill $\blacksquare$

\subsection{Verification of Example~\ref{ex:homophily-adoption}}
\label{app:homophily-example}

Use the networks, abilities, and parameters in
Figure~\ref{fig:composition-homophily}. Denote the initial, intermediate, and final identity profiles by
\[
\begin{aligned}
s^0&=(B,B,B,B,A,A,A,A),\\
s^m&=(A,B,A,B,A,A,A,A),\\
s^A&=(A,\ldots,A).
\end{aligned}
\]
Table~\ref{tab:homophily-verification} reports the degree differences and thresholds for each profile.

\begin{table}[!htbp]
\centering
\begin{threeparttable}
\caption{Threshold comparisons in Example~\ref{ex:homophily-adoption}}
\label{tab:homophily-verification}
\small
\setlength{\tabcolsep}{4pt}
\begin{tabular}{@{}l
  @{\hspace{1em}}r@{\hspace{0.2em}}c@{\hspace{0.2em}}r
  @{\hspace{1em}}r@{\hspace{0.2em}}c@{\hspace{0.2em}}r
  @{\hspace{1em}}r@{\hspace{0.2em}}c@{\hspace{0.2em}}r
  @{\hspace{1em}}r@{\hspace{0.2em}}c@{\hspace{0.2em}}r@{}}
\toprule
& \multicolumn{3}{c}{Original}
& \multicolumn{9}{c}{Rewired}\\
\cmidrule(lr){2-4}\cmidrule(lr){5-13}
Individuals
& \multicolumn{3}{c}{\(s^0\)}
& \multicolumn{3}{c}{\(s^0\)}
& \multicolumn{3}{c}{\(s^m\)}
& \multicolumn{3}{c}{\(s^A\)}\\
\midrule
\(1,3\)
& \(0\)  & \(<\) & \(2.242\)
& \(-2\) & \(>\) & \(-4.022\)
& \(0\)  & \(>\) & \(-10.427\)
& \(4\)  & \(>\) & \(-12.431\)\\
\(2,4\)
& \(-2\) & \(<\) & \(13.785\)
& \(-2\) & \(<\) & \(19.591\)
& \(2\)  & \(>\) & \(-9.362\)
& \(4\)  & \(>\) & \(-10.998\)\\
\(5,7\)
& \(0\) & \(>\) & \(-97.066\)
& \(2\) & \(>\) & \(-102.396\)
& \(2\) & \(>\) & \(-98.610\)
& \(4\) & \(>\) & \(-86.138\)\\
\(6,8\)
& \(1\) & \(>\) & \(-29.654\)
& \(1\) & \(>\) & \(-26.144\)
& \(3\) & \(>\) & \(-35.977\)
& \(3\) & \(>\) & \(-37.232\)\\
\bottomrule
\end{tabular}
\begin{tablenotes}[flushleft]
\footnotesize
\item \textit{Notes:} Each entry compares \(d_{i,A}-d_{i,B}\)
(left) with \(C_i(A,B,s_{-i})\) (right).
Thresholds use the continuation equilibria under the two candidate
identities, holding others' identities fixed, and are rounded to
three decimal places.
\end{tablenotes}
\end{threeparttable}
\end{table}

By Lemma~\ref{lem:threshold}, these comparisons verify the initial and final strict equilibria and, under myopic updating to the observed identity profile, the simultaneous adjustment path \(s^0\to s^m\to s^A\) on the rewired network.

Holding the thresholds of individuals~1 and~3 fixed before rewiring, and those of individuals~2 and~4 fixed at \(s^0\) on the rewired network, gives the following comparisons at their respective switching steps:
\[
\begin{aligned}
i\in\{1,3\}:\quad & d_{i,A}-d_{i,B}=-2<2.242,\\
i\in\{2,4\}:\quad & d_{i,A}-d_{i,B}= 2<19.591.
\end{aligned}
\]
Neither pair would therefore adopt \(A\) under the corresponding
fixed thresholds. At these steps, their reference-group sizes
\((d_{i,A},d_{i,B})\) are \((1,3)\) and \((3,1)\), respectively.
\hfill\(\blacksquare\)

\subsection{Proof of Lemma \ref{lem:homo_actions}}

Under Assumption \ref{assume:homogeneity},
\[
    \psi_i = \frac{w}{1+w\left(\gamma+\alpha 1\{d_{i,I}>0\}\right)}.
\]
Moreover, by construction of $\widehat G_I$,
\[
    \sum_{j\in N_I}\widehat g_{ij,I} = 1\{d_{i,I}>0\}.
\]
Consider the constant centrality vector with entries
\[
    b_{i,I}=\frac{w}{1+\gamma w}.
\]
Substituting into the centrality system implied by equation \eqref{eq:centrality_vector}
gives
\begin{align*}
    \psi_i \left( 1+\alpha\sum_{j\in N_I}\widehat g_{ij,I}b_{j,I}\right)
        &= \frac{w}{1+w\left(\gamma+\alpha 1\{d_{i,I}>0\}\right)}\left(1+\alpha 1\{d_{i,I}>0\}\frac{w}{1+\gamma w} \right) \\
            &= \frac{w}{1+\gamma w}.
\end{align*}
Hence this vector satisfies the centrality system. By uniqueness from Lemma \ref{lem:action_equ},
\[
    b_{i,I} = \frac{w}{1+\gamma w} \qquad \text{for all } i\in N_I.
\]
Equation \eqref{eq:equilibrium_x} therefore implies
\[
    x_{i,I}^* = (1+\gamma v_I)b_{i,I} = (1+\gamma v_I)\frac{w}{1+\gamma w}.
\]
Since all individuals choosing identity $I$ take the same equilibrium action, $\bar{x}_{i,I}=x_{i,I}^*$ whenever $d_{i,I}>0$; when $d_{i,I}=0$, the peer-conformity term is zero by the indicator. Hence
\[
    \widetilde V_i(I,s_{-i}) = \mu_I +x_{i,I}^* - \frac{(x_{i,I}^*)^2}{2w} - \frac{\gamma}{2}(x_{i,I}^*-v_I)^2,
\]
which is independent of both $i$ and $s_{-i}$. Consequently,
$C_i(I,J,s_{-i})$ is also identical across individuals and independent of $s_{-i}$, and we denote this common value by $C(I,J)$. \hfill $\blacksquare$

\subsection{Proof of Proposition \ref{prop:homogeneous_coexistence}}

By Lemma \ref{lem:homo_actions}, pairwise costs are common across individuals and independent of the identity profile. Moreover, \(C(B,A)=-C(A,B)\). Hence, by Theorem \ref{prop:equilibria_pure_strategy}, an individual choosing \(B\) must satisfy \(d_i(N_B)-d_i(N_A)\ge -C(A,B)\), whereas an individual choosing \(A\) must satisfy \(d_i(N_A)-d_i(N_B)\ge C(A,B)\). In an all-\(B\) profile, \(N_B=N\) and \(N_A=\varnothing\), so the first condition reduces to \(d_i\ge -C(A,B) \) for every \(i\in N\), proving part~\emph{(i)}.

For coexistence, let \(S=N_B\), so that \(S^C=N_A\). By Definition \ref{def:k(S)}, the two conditions become \( k_i(S)\ge -C(A,B)\) for all $i\in S$, and \(k_i(S^C)\ge C(A,B)\) for all $i\in S^C$, which proves part~\emph{(ii)}. \hfill $\blacksquare$

\subsection{Proof of Corollary \ref{coroll:density}}
\label{app:proof-density}
Part~\emph{(i)} follows immediately from Proposition~\ref{prop:homogeneous_coexistence} part~\emph{(i)}.

For part~\emph{(ii)}, Proposition~\ref{prop:homogeneous_coexistence} implies that \(k_i(S)\ge -C(A,B)>0\) for every \(i\in S\). Under proportional densification,
\[
k_i'(S)=\lambda k_i(S)>k_i(S),
\]
so every individual initially choosing \(B\) continues to find \(B\) a best response. Starting from \(S\), add any individuals who prefer \(B\) in the denser network. Expanding the set of \(B\)-adopters weakly increases \(k_i\) with respect to that set for every existing \(B\)-adopter, so no individual already added acquires an incentive to leave \(B\). Because the population is finite, this monotone procedure terminates at a set \(S'\supseteq S\) at which no further profitable switch to \(B\) remains; the resulting profile is a pure-strategy equilibrium. \hfill\(\blacksquare\)

\subsection{Proof of Proposition \ref{prop:diffusion}}
\label{app:proof-diffusion}
Starting from the all-\(B\) equilibrium, an individual switching alone to \(A\) gives up all same-identity neighbors. By Lemma~\ref{lem:threshold}, individual \(i\) therefore strictly adopts \(A\) in the first round if and only if
\[
d_i<-C'(A,B),
\]
where equality is excluded by assumption. Hence diffusion begins if and only if \(\min_i d_i<-C'(A,B)\), proving part~\emph{(i)}.

For part~\emph{(ii)}, consider a nonempty set \(S\) of individuals choosing \(B\). If
\[
k_i(S)\ge -C'(A,B)\qquad\text{for every }i\in S,
\]
then every member of \(S\) prefers to remain with \(B\) even when every individual outside \(S\) has adopted \(A\), so full diffusion is impossible. Conversely, if diffusion terminates with a nonempty set \(S\) choosing \(B\), the terminal profile is a best-response fixed point and the same inequalities must hold for all \(i\in S\).

Finally, once an individual adopts \(A\), subsequent switches from \(B\) to \(A\) weakly increase that individual's relative social support for \(A\), so they never switch back. Since the population is finite, the process terminates after finitely many rounds. \hfill\(\blacksquare\)

\subsection{Proof of Proposition \ref{prop:welfare_diffusion}}
\label{app:proof-welfare-diffusion}
Relative to the original all-\(B\) equilibrium, each adopter obtains the core-value gain \(-\beta C'(A,B)\). Every link between \(N_A'\) and \(N_B'\), however, ceases to provide the same-identity benefit to both endpoints and therefore reduces aggregate welfare by \(2\beta\). Hence
\[
W'-W=\beta\bigl[-C'(A,B)|N_A'|-2\ell'\bigr],
\]
which establishes \eqref{eq:homogeneous-welfare-change} and \eqref{eq:welfare_loss}.

Individuals retaining \(B\) receive no core-value gain and lose the same-identity benefit on every link to \(N_A'\). They are therefore weakly worse off, with equality if and only if they have no neighbors in \(N_A'\). An adopter \(i\in N_A'\) gains \(-\beta C'(A,B)\) from choosing \(A\), but loses \(\beta\) for each neighbor retaining \(B\). Their change in utility relative to the original all-\(B\) equilibrium is therefore
\[
\beta\bigl[-C'(A,B)-d_i(N_B')\bigr].
\]
Thus, the adopter is strictly worse off if and only if \eqref{eq:individual_welfare_loss} holds.

Finally, every first-round adopter satisfies \(d_i<-C'(A,B)\). Since \(d_i(N_B')\le d_i\), it follows that \(d_i(N_B')<-C'(A,B)\), so every first-round adopter is strictly better off in the terminal equilibrium than in the original all-\(B\) equilibrium. \hfill\(\blacksquare\)

\subsection{Algorithm for identity diffusion} \label{A:algorithm}

The following procedure is analogous to the \(q\)-core algorithm of \cite{gagnon-goyal}, but applies to degree differences rather than an absolute number of links.

\begin{algorithm}[Degree-difference diffusion] \label{algorithm}
Let \(B^0=N\) denote the initial set of individuals choosing identity \(B\). For \(t=0,1,\ldots\), define
    \[
        B^{t+1} = \left\{i\in B^t: k_i(B^t)\geq -C'(A,B)\right\}
    \]
All individuals in \(B^t\setminus B^{t+1}\) adopt identity \(A\). Repeat until \(B^{t+1}=B^t\).
\end{algorithm}

The sequence \(B^0\supseteq B^1\supseteq\cdots\) is decreasing and therefore terminates after finitely many iterations. Moreover, once an individual adopts \(A\), subsequent adoption of \(A\) by others can only increase their relative social support for \(A\). Identity changes are therefore irreversible along this adjustment path. The terminal set is the maximal group of individuals that can remain sufficiently embedded in \(B\) to resist the new core-value advantage of \(A\); full diffusion occurs if and only if this set is empty. \hfill$\blacksquare$

\subsection{Verification of the designed-assignment example}
\label{A:carrell}

Use the parameters and group compositions in
Section~\ref{subsec:designed_peers}. Table~\ref{tab:carrell-thresholds} reports all distinct comparisons along the specified adjustment path and for composition \((4,3,1)\).

\begin{table}[!htbp]
\centering
\begin{threeparttable}
\caption{Threshold comparisons under alternative assignments}
\label{tab:carrell-thresholds}
\small
\begin{tabular}{@{}lc
  @{\hspace{1em}}r@{\hspace{0.2em}}c@{\hspace{0.2em}}r
  @{\hspace{1em}}r@{\hspace{0.2em}}c@{\hspace{0.2em}}r
  @{\hspace{1em}}r@{\hspace{0.2em}}c@{\hspace{0.2em}}r@{}}
\toprule
Composition & Profile
& \multicolumn{3}{c}{Low}
& \multicolumn{3}{c}{Middle}
& \multicolumn{3}{c}{High}\\
\midrule
\((4,4,0)\) & \((H,H,\text{--})\)
& \(7\) & \(>\) & \(6.832\)
& \(7\) & \(>\) & \(1.604\)
& \multicolumn{3}{c}{--}\\
\((0,4,4)\) & \((\text{--},H,H)\)
& \multicolumn{3}{c}{--}
& \(7\) & \(>\) & \(-5.157\)
& \(7\) & \(>\) & \(-5.607\)\\
\((4,3,1)\) & \((H,H,H)\)
& \(7\) & \(>\) & \(6.886\)
& \(7\) & \(>\) & \(1.553\)
& \(7\) & \(>\) & \(1.956\)\\
\addlinespace
\((4,0,4)\) & \((H,\text{--},H)\)
& \(7\) & \(<\) & \(7.053\)
& \multicolumn{3}{c}{--}
& \(7\) & \(>\) & \(1.794\)\\
\((4,0,4)\) & \((L,\text{--},H)\)
& \(1\) & \(<\) & \(20.030\)
& \multicolumn{3}{c}{--}
& \(-1\) & \(>\) & \(-14.647\)\\
\((0,8,0)\) & \((\text{--},H,\text{--})\)
& \multicolumn{3}{c}{--}
& \(7\) & \(>\) & \(-5.167\)
& \multicolumn{3}{c}{--}\\
\bottomrule
\end{tabular}
\begin{tablenotes}[flushleft]
\footnotesize
\item \textit{Notes:} Compositions and profiles list low, middle, and
high types. Each entry compares \(d_{i,H}-d_{i,L}\) (left) with
\(C_i(H,L,s_{-i})\) (right). Thresholds use the action equilibria under
each candidate identity, holding all other identities fixed, and are
rounded to three decimal places.
\end{tablenotes}
\end{threeparttable}
\end{table}

By Lemma~\ref{lem:threshold}, the comparisons verify the strict
benchmark equilibria and the strict all-\(H\) equilibrium at
\((4,3,1)\). Under the designed assignment, one simultaneous
best-response round changes only low types' identity; the
resulting profiles are strict equilibria. Holding thresholds
at their benchmark values (the first two rows) instead
preserves the all-\(H\) profile: every benchmark threshold
is below the unchanged link difference of seven. Actions
in Table~\ref{tab:carrell} follow from
equation~\eqref{eq:equilibrium_x}.
\hfill\(\blacksquare\)


\newpage
\bibliographystyle{aea}
\bibliography{bibidentity}

\end{document}